\documentclass[12pt, a4paper, preprint]{article}
\pdfoutput=1
\usepackage{jheppub}
\usepackage{bm} 
\usepackage{tikz}

\def\bea{\begin{eqnarray}}
\def\eea{\end{eqnarray}}
\def\beas{\begin{eqnarray*}}
\def\eeas{\end{eqnarray*}}

\title{Emergent Unitarity from the Amplituhedron}
\author{Akshay Yelleshpur Srikant}
\affiliation{Department of Physics, Princeton University, NJ, USA}
\date{}
\abstract{We present a proof of perturbative unitarity for planar $\mathcal{N}=4$ SYM, following from the geometry of the amplituhedron. This proof is valid for amplitudes of arbitrary multiplicity $n$, loop order $L$ and MHV degree $k$.}
\begin{document}
\maketitle
\newpage
\section{Introduction}
Unitarity is at the heart of the traditional, Feynman diagrammatic approach to calculating scattering amplitudes. It is built into the framework of quantum field theory. Modern on-shell methods provide an alternative way to calculate scattering amplitudes. While they eschew Lagrangians, gauge symmetries, virtual particles and other redundancies associated with the traditional formalism of QFT, unitarity remains a central principle that needs to be imposed. It has allowed the construction of loop amplitudes from tree amplitudes via generalized Unitarity methods \citep{unitarity1, unitarity2, unitarity3, unitarity4, unitarity5} and the development of loop level BCFW recursion relations \cite{bcfw1,bcfw2}. These on-shell methods have been particularly fruitful in planar $\mathcal{N}=4$ SYM and led to the development of the on-shell diagrams in \cite{onshelldiagrams} and the discovery of the underlying Grassmannian structure. Locality and unitarity seemed to be the guiding principles which dictated how the on-shell diagrams glued together to yield the amplitude. The discovery of the amplituhedron in \cite{amplituhedron}, \cite{intotheamplituhedron} revealed the deeper principles behind this process - positive geometry. Positivity dictated how the on-shell diagrams were to be glued together. The resulting scattering amplitudes were miraculously local and unitary!  \\

This discovery of the amplituhedron was inspired by the polytope structure of the six point NMHV scattering amplitude, first elucidated in \citep{hodges} and expanded upon in \cite{polytopes}. This motivated the original definition of the amplituhedron which was analogous to the definition of the interior of a polygon. The tree amplituhedron $\mathcal{A}_{n,k,0}$ was defined as the span of $k$ planes $Y^I_\alpha$, living in $(k+4)$ dimensions. Here $I = \lbrace 1, \dots , k+4\rbrace$ and $\alpha = \lbrace 1, \dots k\rbrace$. 
\bea
\label{eq:oldamp}
Y^I_\alpha = C_{\alpha a}\mathcal{Z}_a^I
\eea
where $\mathcal{Z}_a^I\, (a = 1, \dots n)$ are positive external data in $(k+4)$ dimensions. In this context, positivity refers to the conditions $det \left\lbrace \mathcal{Z}_{a_1}, \dots \mathcal{Z}_{a_{k+4}}\right\rbrace \equiv \langle \mathcal{Z}_{a_1}\dots \mathcal{Z}_{a_{k+4}}\rangle > 0$ if $a_1< \dots <a_{k+4}$ and $C_{\alpha a} \in G_+(k,n)$. $G_+(k,n)$ is the positive Grassmannian defined as the set of all $k\times n $ matrices with ordered, positive $k\times k$ minors. For more details on the properties of the positive Grassmannian, see \cite{onshelldiagrams, grassmannian1, grassmannian2, grassmannian3} and the references therein. The scattering amplitude can be related to the differential form with logarithmic singularities on the boundaries of the amplituhedron. The exact relation along with the extension of eq.(\ref{eq:oldamp}) to loop level can be found in \cite{amplituhedron}. 
\\

The amplituhedron thus replaced the principles of unitarity and locality by a central tenant of positivity. Tree level locality emerges as a simple consequence of the boundary structure of the amplituhedron, which in turn is dictated by positivity. The emergence of unitarity is more obscure. It is reflected in the factorization of the geometry on approaching certain boundaries. This was proved for $\mathcal{A}_{4,0,L}$ in \cite{intotheamplituhedron}. The extension of this proof to amplitudes with arbitrary multiplicity using (\ref{eq:oldamp}) is cumbersome and requires the use of the topological definition of the amplituhedron introduced in \cite{binarycode}. In the following section, we review this definition in some detail along with some properties of scattering amplitudes relevant to this paper. We also expound the relation between the amplituhedron and scattering amplitudes. The rest of the paper is structured as follows. In Section [\ref{sec:4pt}], we present a proof of unitarity of scattering amplitudes for 4 point amplitudes of planar $\mathcal{N}=4$ SYM, using the topological definition of the amplituhedron. This serves as a warm up to Section [\ref{sec:MHVproof}] in which we provide a proof which is valid for MHV amplitudes of any multiplicity. Finally, in Section \ref{sec:higherk} we show how the proof of the previous section can be extended to deal with the complexity of higher $k$ sectors. 

\section{Review of the topological definition of $\mathcal{A}_{n,k,L}$}
\label{sec:review}
The scattering amplitudes extracted from the amplituhedron defined as in ($\ref{eq:oldamp}$), using the procedure outlined in \cite{amplituhedron}, reproduce the Grassmannian integral form of scattering amplitudes presented in \cite{grassmannian1, grassmannian2, grassmannian3, grassmannian4, grassmannian5}. These necessarily involve the auxiliary variables $C_{\alpha a}$. In contrast, the topological definition of the amplituhedron can be stated entirely in terms of the $4D$ momentum twistors (first introduced in \cite{hodges}). Consequently, this yields amplitudes that can be thought of as differential forms on the space of momentum twistors. In this section, we will review the basic concepts involved in the topological definition of the amplituhedron. We begin with a review of momentum twistors and their connection to momenta in Section [\ref{sec:momtwist}] and proceed to the topological definition of the amplituhedron in Section [\ref{sec:top def}]. We then explain how amplitudes are extracted from the amplituhedron in Section [\ref{sec:amps}] and finally, in Section [\ref{sec:Opticaltheorem}], we set up the statement of the optical theorem in the language of momentum twistors. This is the statement we will prove in the main body of the paper.  
\subsection{Momentum twistors} 
\label{sec:momtwist}
\begin{figure}[htb!]
\label{fig:momtwistgeometry}
\begin{center}
\includegraphics[scale=.6]{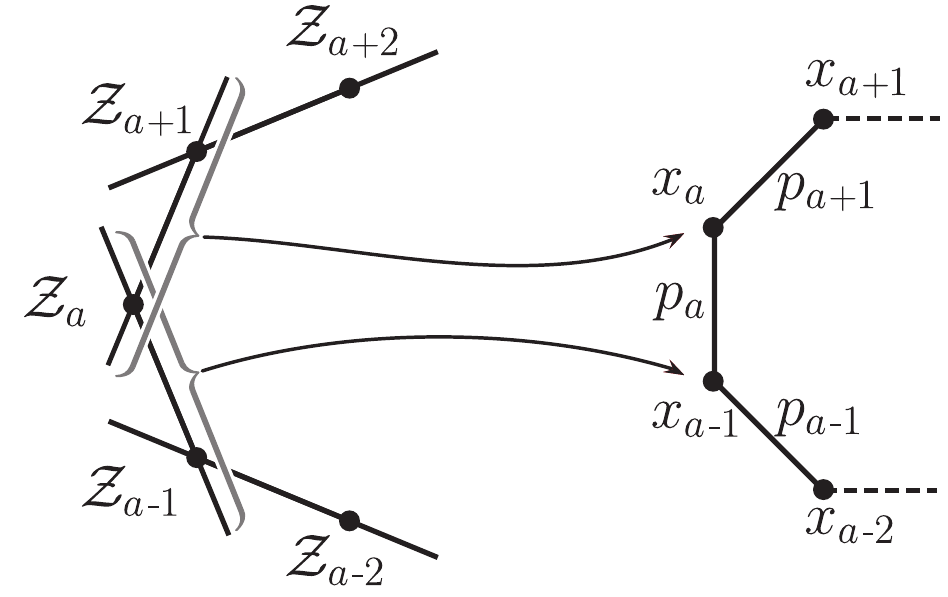}
\caption{A representation of the relationship between momenta and momentum twistors, taken from \cite{LocalIntegrands}.}
\end{center}
\end{figure}

Momentum twistor space is the projective space $\mathbb{CP}^3$. A connection to physical momenta can be made by writing them in the coordinates of an embedding $\mathbb{C}^4$ as $Z_a = \left(\lambda_{a\,\alpha}, \mu_a^{\dot{\alpha}}\right)$. Here $(\lambda, \tilde{\lambda})$ are the spinor helicity variables which trivialize the on-shell condition. 
\beas
&& p_{a\,\alpha \dot{\alpha}} \equiv \lambda_{a\,\alpha} \tilde{\lambda}_{a\, \dot{\alpha}} \implies \, \, p_a^2 = \text{det} (\lambda_a, \lambda_a)\, \text{det} (\tilde{\lambda}_a,\tilde{\lambda}_a) = 0
\eeas 
$\mu_a^{\dot{\alpha}} = x_a^{\alpha \dot{\alpha}}\lambda_{a\alpha} $ where the dual momenta $x_a$ are defined via $p_a = x_a - x_{a-1}$ and trivialize conservation of momentum. Thus the point $x_a$ in dual momentum space is associated to a line in momentum twistor space. Scattering amplitudes in $\mathcal{N}=4$ SYM involve momenta $p_a$ which are null $(p_a^2 = 0)$ and are conserved $(\sum_a p_a = 0)$. Momentum twistors are ideally suited to describe the momenta involved in these amplitudes because they trivialize both these constraints. Figure \ref{fig:momtwistgeometry}  summarizes the point-line correspondence between points in dual momentum space and lines in momentum twistor space. 
Thus all the points of the form
\bea
\label{eq:momtwist}
Z_a = \left( \lambda_{a\,\alpha}, \,x_a^{\alpha \dot{\alpha}}\lambda_{a\alpha} \right)
\eea 
are associated to the momentum $p_a$. Note that these $Z_a$ are different from the calligraphic $\mathcal{Z}_a$ used in  \ref{eq:oldamp}(the connection between the two is that $Z_a$ are obtained by projecting the $\mathcal{Z}_a$ through the $k-$plane $Y$ in \ref{eq:oldamp}). Thus, a set of on-shell momenta $\left\lbrace p_1, \dots p_n\right\rbrace$ satisfying $\sum_{a=1}^n p_a=0$ can be represented by an ordered set of momentum twistors $\left\lbrace Z_1, \dots Z_n \right\rbrace$. Each line $Z_aZ_{a+1}$ corresponds to the point $x_a$ in dual momentum space as shown in Fig \ref{fig:momtwistgeometry}.\\
 
Each loop momentum $\ell_a$ can also be associated to a line in momentum twistor space. We denote these lines by $(AB)_a$, where $A$ and $B$ are any representative points. This helps us distinguish loop momenta from other external momenta. We can express Lorentz invariants in terms of determinants of momentum twistors using the relation
\bea
\label{eq: 4bracket}
(x_a - x_b)^2 = \frac{\langle a-1ab-1b\rangle}{\langle a-1a\rangle \langle b-1b\rangle} 
\eea
with $\langle abcd \rangle = \text{det }\left\lbrace Z_a, Z_b, Z_c, Z_d \right\rbrace$ and $\langle ab\rangle = \text{det }\left\lbrace I_{\infty}, Z_a, Z_b \right\rbrace$ where $I_\infty$ is the infinity twistor \citep{LocalIntegrands, grassmannian2}. Finally, we connect the invariants involving the $\mathcal{Z}_a^I$ with the four bracket via 
\beas
\langle Z_a Z_b Z_c Z_d\rangle = \epsilon_{I_1 \dots I_{k+4}} Y_{1}^{I_1} \dots Y_k^{I_k}\mathcal{Z}_a^{I_{k+1}}\mathcal{Z}_b^{I_{k+2}}\mathcal{Z}_c^{I_{k+3}}\mathcal{Z}_d^{I_{k+4}}
\eeas
We will utilize this connection later in Section [\ref{sec:higherkmutual}].
\subsection{Topological definition}
\label{sec:top def}
The amplituhedron $\mathcal{A}_{n,k,L}$ is a region in momentum twistor space which can be cut out by inequalities. The region depends on the integers $n, k$ and $L$ which specify the $n-$point, N$^k$MHV amplitude. $n$ is the number external legs of the amplitude and correspondingly the number of momentum twistors which are involved in the definition of the amplituhedron. We denote these by $\left\lbrace Z_1, \dots Z_n \right\rbrace$. $L$ is the number of loops and we have the lines, $(AB)_1, \dots \dots (AB)_L$ corresponding to the $L$ loop momenta $\ell_1, \dots \ell_L$. $k$ appears below in the inequalities that define $\mathcal{A}_{n,k,L}$. 
\subsubsection{Tree level conditions} 
\label{sec:treelevelconditions}
The first set of conditions that define the amplituhedron involve only the external momentum twistors $Z_a$ and we refer to these as the ``tree-level" conditions. They are listed below along with some comments about each condition.\\
\begin{itemize}
\item The external data must satisfy the following positivity conditions.
\bea
\label{eq:kamplituhedrontree1}
&&\langle ii+1jj+1 \rangle > 0 \quad i=1, \dots n
\eea
We adopt an ordering $(1, \dots n)$ in all definitions. We must also define a twisted cyclic symmetry for this ordering with 
\bea
\label{eq:twistedcyclic}
Z_{n+i} \equiv (-1)^{k-1} Z_i
\eea
This definition is required to ensure that $\langle ii+1n1\rangle >0$ for odd $k$ and $\langle ii+1n1\rangle <0$ for even $k$. We will see below that this is crucial to obtain the right number of sign flips. 
\item We require that the sequence
\bea
\label{eq:kamplituhedrontree2}
&&S^{\text{tree}}:\left\lbrace \langle 1234 \rangle, \langle 1235\rangle \dots \langle 123n\rangle \right\rbrace \text{ has $k$ sign flips.}
\eea
Note that the ordering $(1, \dots n)$ is crucial for the above condition to make sense. Using (\ref{eq:kamplituhedrontree1}), (\ref{eq:kamplituhedrontree2}) and the reasoning in Appendix [\ref{sec:app1}], we can conclude that all sequences of the form 
\beas
\left\lbrace \langle ii+1i+2i+3\rangle, \dots \langle ii+1i+2n\rangle, (-1)^{k-1} \langle ii+1i+21\rangle, \dots \langle ii+1i+2i-1\rangle (-1)^{k-1} \right\rbrace
\eeas
with $i=1, \dots n$ have $k$ sign flips. The use of (\ref{eq:twistedcyclic}) is crucial in arriving at this conclusion. Since all the sequences $\left\lbrace \langle ii+1i+2j\rangle \right\rbrace_{j=i+3}^{j=i-1}$ have the same number of flips (with the appropriate twisted cyclic symmetry factors), we can use any of them in place of the sequence $\left\lbrace \langle 123i \rangle \right\rbrace_{i=4}^{i=n}$. In the rest of the paper, we will the sequence which is most convenient to the situation.  
\end{itemize}
\subsubsection{Loop level conditions}
\label{sec:looplevelconditions}
The next set of conditions involve both the external data and the loops $(AB)_a$ and we refer to these as ``loop level" conditions.  
\begin{itemize}
\item Each loop $(AB)_a$ must satisfy a positivity condition analogous to (\ref{eq:kamplituhedrontree1})
\bea
\label{eq:kamplituhedronloop1}
&&\langle (AB)_aii+1 \rangle >0 \quad i=1, \dots n
\eea
Note that we must include the twisted cyclic symmetry factor $(-1)^{k-1}$ here as well. Once again, this implies that $\langle (AB)_a n1 \rangle >0$ for odd $k$ and $\langle (AB)_a 1n\rangle >0$ for even $k$. 
\item We require that sequence  
\bea 
\label{eq:kamplituhedronloop2}
&& S^{\text{loop}}: \left\lbrace \langle (AB)_a 12\rangle, \langle (AB)_a 13\rangle, \dots \langle (AB)_a 1n\rangle \right\rbrace
\text{has $k+2$ flips.}
\eea
Following a line of reasoning similar to that in $\ref{sec:treelevelconditions}$, we can show that all sequences of the form 
\beas
\left\lbrace \langle (AB)_a ii+1\rangle, \dots \langle (AB)_ain\rangle, \langle (AB)_ai1\rangle (-1)^{k-1}, \dots \langle (AB)_aii-1\rangle (-1)^{k-1}\right\rbrace
\eeas 
with $i=1, \dots n$ have the same number of sign flips. We will make use of these sequences as convenient in the rest of the paper.
\end{itemize}
\subsubsection{Mutual positivity condition} 
\label{sec:mutualpositivitycondition}
The final condition is a relation involving multiple loop momenta $(AB)_a$. We must have
\bea
\label{eq:mutualpositivtycondition}
\langle (AB)_a(AB)_b \rangle >0 \qquad \forall a,b = \left\lbrace 1, \dots L\right\rbrace
\eea
For multi loop amplitudes, the conditions above amount to demanding that each loop $(AB)_a$ is in the one-loop amplituhedron (i.e. it satisfies conditions $(\ref{eq:kamplituhedronloop1})$ and $(\ref{eq:kamplituhedronloop2})$) and also the mutual positivity condition (\ref{eq:mutualpositivtycondition}). Finding a solution to all these inequalities is tantamount to computing the $n-$point N$^k$MHV amplitude. The complexity of solving the mutual positivity condition shows up even in the simplest case of $n=4$. Indeed, its solution is at the heart of the four point problem, as explained in \cite{intotheamplituhedron}. \\

The topological definition is well suited to exploring cuts of amplitudes (which correspond to saturating some of the inequalities in (\ref{eq:kamplituhedrontree1}) - (\ref{eq:mutualpositivtycondition}) by setting them to be equal to zero). This formalism has been exploited to investigate the structure some cuts of amplitudes that are inaccessible by any other means. The results of some classes of these ``deep" cuts are obtained to all loop orders in \cite{mypaper1,mypaper2}.
\subsection{Amplitudes and integrands as Canonical Forms} 
\label{sec:amps}
The inequalities (\ref{eq:kamplituhedrontree1}) - (\ref{eq:mutualpositivtycondition}) define a region in the space of momentum twistors. The goal of the amplituhedron program is to be able to obtain the amplitude from purely geometric considerations. More precisely, we can obtain the tree level amplitude and the loop level integrand for planar $\mathcal{N}=4$ SYM. In contrast to generic quantum field theories, the planar integrand in $\mathcal{N} = 4$ SYM is a well defined, rational function as shown in \cite{unitarity6, bcfw2}. The conjecture here is that the Canonical form associated to the amplituhedron is the loop integrand. The Canonical form associated to a region is the differential form with logarithmic singularities on all the boundaries of that region. For more details on Canonical forms, their properties and precise definitions, see \cite{canonicalforms}. The discovery of amplituhedron-like geometric structures (for e.g. \cite{othergeometries1, othergeometries2, othergeometries3, othergeometries4}) in other theories lends further support to the idea that amplitudes can be thought of as differential forms on kinematic spaces. Some consequences of this are explored in \cite{ampsasdiff}. It is interesting to note that a topological definition of the amplituhedron has been found directly in momentum space \cite{amplituhedronmomentum}. This allows for the possibility of expressing $\mathcal{N}=4$ SYM amplitudes as differential forms in momentum space rather than momentum twistor space. \\

It is illustrative to show the calculation of the canonical form for the simple case of $\mathcal{A}_{4,0,1}$. This canonical form should be the 4-point, one-loop MHV integrand. The defining inequalities are 
\bea
\label{eq:4ptinequalities}
&&\underline{\text{Tree Level}: }\,\,\, \langle 1234\rangle >0\\
&&\nonumber\underline{\text{Loop Level}: }\, \langle AB12\rangle >0, \,\, \langle AB23\rangle > 0, \,\, \langle AB34\rangle > 0, \,\, \langle AB14\rangle > 0,\\
&&\nonumber\hspace{2.35cm}\langle AB13 \rangle <0, \,\, \langle AB24\rangle <0 
\eea  
Since $A, B \in \mathbb{C}^4$, we can expand these in a basis consisting of $\left\lbrace Z_1, Z_2, Z_3, Z_4 \right\rbrace$. However, $A, B$ are arbitrary points on the line $AB$ which corresponds to the loop momentum. Since any linear combination $(A',B')$ of the points $A$ and $B$ is also on the same line, there is a $GL(2)$ redundancy in the choice of $A$ and $B$. Fixing this redundancy, we arrive at the following parametrization. 
\beas
A = Z_1 + \alpha_1 Z_2 + \alpha_2 Z_3 \qquad B = -Z_1+ \beta_1 Z_3  + \beta_2 Z_4
\eeas
The solution to the inequalities in (\ref{eq:4ptinequalities}) is
\beas
\alpha_1 \, > \, 0\, \qquad \alpha_2 \, > \, 0 \, \qquad \beta_1 \, > \, 0\, \qquad \beta_2 \, >0 \,  
\eeas
The boundaries are located at $\alpha_1 = \alpha_2 = \beta_1 = \beta_2 = 0$ and the differential form with logarithmic singularities on all the boundaries is just
\beas
\frac{d\alpha_1}{\alpha_1}\frac{d\alpha_2}{\alpha_2}\frac{d\beta_1}{\beta_1}\frac{d\beta_2}{\beta_2} = \frac{\langle ABd^2A\rangle \langle ABd^2B\rangle}{\text{Vol}(GL(2))}\frac{\langle 1234\rangle^2}{\langle AB12\rangle \langle AB23\rangle \langle AB34\rangle \langle AB14\rangle}
\eeas
This is the integrand for the 1-loop four point MHV amplitude as conjectured. \\

At higher points, the situation is more complicated cases. There are multiple ways in which the sequence $S^{\text{loop}}$ (\ref{eq:kamplituhedronloop2}) can have $k+2$ sign flips. It is useful to triangulate the complete region by enumerating all such patterns. This procedure works extremely well for $n-$point MHV amplitudes and is fleshed out in Section [7] of \cite{binarycode}. Specifically, if we parametrize
\bea
\label{eq:Kermit}
A_a = Z_{1} + \alpha_1 Z_i + \alpha_2 Z_{i+1} \quad B_a = -Z_1 + \beta_1 Z_j + \beta_2 Z_{j+1} 
\eea 
with $\alpha_i \, > \, 0, \, \beta_i \, > \, 0$, we have $\frac{\langle (AB)_a1i \rangle}{\langle (AB)_a1i+1\rangle} = -\frac{\alpha_1}{\alpha_2} $ and $\frac{\langle (AB)_a1j \rangle}{\langle (AB)_a1j+1\rangle} = -\frac{\beta_1}{\beta_2} $. The sequence $\left\lbrace \, \langle (AB)_a1i\rangle \right\rbrace$ has two sign flips, one occurring between $\langle (AB)_a1i \rangle $ and $\langle (AB)_a1i+1\rangle$ and the second between $\langle (AB)_a1j \rangle $ and $\langle (AB)_a1j+1\rangle$. Summing over all $i<j = \left\lbrace 1, \dots n\right\rbrace$ covers all the positions of the flips. The canonical form for this region is the integrand of the $n - $ point MHV amplitude.
\beas
\sum_{i<j} \langle ABd^2A\rangle \langle ABd^2B\rangle \frac{\langle AB(1ii+1 \cap 1jj+1)\rangle^2}{\langle AB1i\rangle \langle AB1i+1\rangle \langle ABii+1\rangle \langle AB1j\rangle \langle AB1j+1\rangle \langle ABjj+1\rangle }
\eeas
For another derivation of this integrand, please refer to \cite{bcfw2}. \\

Finally, an important property of these forms is that they are all projectively well defined. They are invariant under the re-scaling $Z_i \rightarrow t_i Z_i$ of each external leg. We will make use of this property in Section [\ref{sec:higherk}].

\subsection{Unitarity and the Optical theorem}
\label{sec:Opticaltheorem}
The relationship between the singularity structure of scattering amplitudes and unitarity has been the subject of a lot of work. \cite{unitarity1, unitarity2, unitarity3, unitarity4, unitarity5, unitarity6, unitarity7, unitarity8, unitarity9}. It is well known that the branch cut structure of amplitudes is intimately tied to perturbative unitarity. This is encapsulated in the optical theorem which related the discontinuity across a double cut to the product of lower loop amplitudes. \\

The presence of branch points in loop amplitudes is due to the pole structure of the integrand. This is governed by the boundary structure of the amplituhedron. The structure of boundaries and their relation to branch points has been studied extensively in \cite{boundary1, boundary2, boundary3, boundary4, boundary5, boundary6, boundary7, boundary8}. The discontinuity across a branch cut is calculated by the residue on an appropriate boundary of the amplituhedron. The optical theorem thus translates into a statement about the factorization of the residue on this boundary. We expect this factorization to emerge as a consequence of the positive geometry. \\
\begin{figure}[htb!]
\begin{center} 
\label{fig:MHVcut}
\includegraphics[scale=.4]{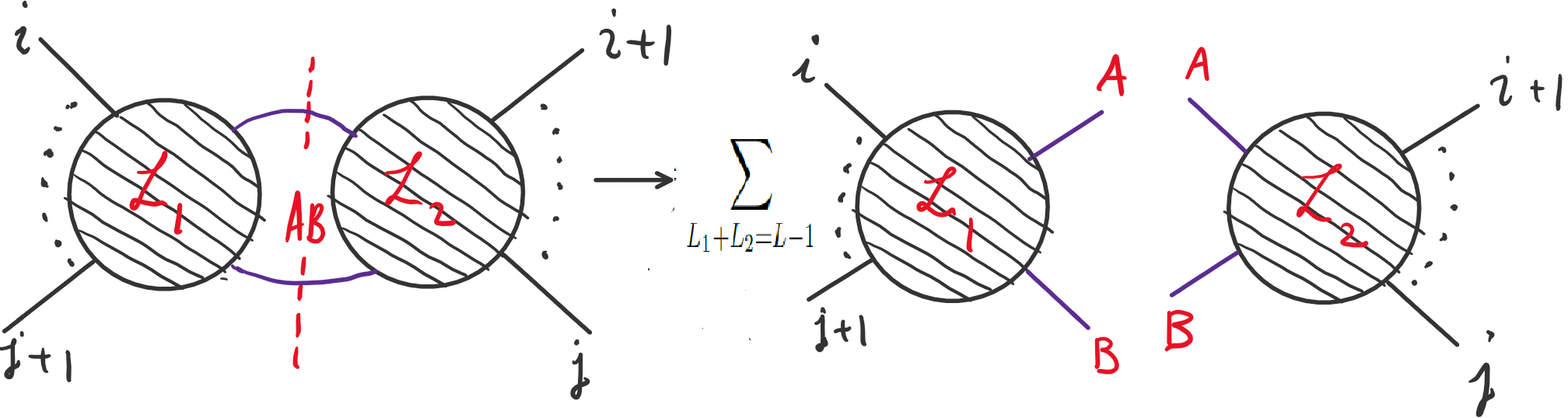}
\caption{Structure of a unitarity cut of MHV amplitudes. The loop $(AB)$ is cut and the residue factorizes as shown.}
\end{center}
\end{figure}
\\\\
Let us begin by rewriting the optical theorem, specifically for $\mathcal{N}=4$ SYM in the language of momentum twistors. For now, we will focus on MHV amplitudes. We are interested in the case where one of the loops, $AB$, cuts the lines $ii+1$ and $jj+1$ and all other loops (which we denote by $(AB)_a$) remain uncut. Thus we are calculating the residue of the $n-point$ MHV amplitude on the cut $\langle ABii+1\rangle = \langle ABjj+1\rangle = 0$. It is convenient to parametrize $AB$ as 
\bea
\label{eq:cut loop}
A = Z_i + x Z_{i+1} + w_1 Z_{\star} \qquad B = y Z_j + Z_{j+1} + w_2 Z_{\star}
\eea 
where $Z_{\star}$ is an arbitrary reference twistor. The terms in the $L$-loop integrand which contribute to this cut (which has the required poles) can be written as
\beas
\mathcal{M}_n^L = \frac{\langle ABd^2A \rangle \langle ABd^2B\rangle}{\langle ABii+1\rangle \langle ABjj+1\rangle} \prod_{a=1}^{L-1} \langle(AB)_ad^2A_a\rangle \langle(AB)_ad^2B_a\rangle f(x,y,w_1,w_2, (AB)_a)
\eeas
The dependence of $f$ on the external twistors has been suppressed. The residue of $\mathcal{M}_n^L$ on the cut $\langle ABii+1\rangle = \langle ABjj+1\rangle = 0$ is 
\bea
\label{eq:cut}
Res_{w_1=w_2=0} \, \mathcal{M}_n^L = \frac{dx\, dy}{g(x,y)}\prod_{a=1}^{L-1} \langle(AB)_ad^2A_a\rangle \langle(AB)_ad^2B_a\rangle f(x,y,0,0 ,(AB)_a)
\eea
where $g(x,y)$ is a Jacobian which is irrelevant to our purposes. Unitarity predicts that the function $f(x,y,0,0,(AB)_a)$ is related to lower point amplitudes (see figure \ref{fig:MHVcut}) and is of the form
\bea
\label{eq:Unitarity}
\hspace{-1cm}f(x,y,0,0,(AB)_a) =\sum_{L_1+L_2 = L-1} \mathcal{M}_{\mathcal{L}}^{L_1}(Z_{j+1}, \dots Z_i,A,B)  \mathcal{M}_{\mathcal{R}}^{L_2}(B, A, Z_{i+1}, \dots , Z_j)
\eea
We will show that this structure emerges from the geometry of the amplituhedron. We will first present a proof for the four point case. This is just a rewriting of the proof found in \cite{intotheamplituhedron} using the topological definition. This proof will then admit a generalization to amplitudes of higher multiplicity.

\section{Proof for 4 point Amplitudes}
\label{sec:4pt}
\begin{figure}[htb!]
\begin{center}
\label{fig:4pt}
\includegraphics[scale=.3]{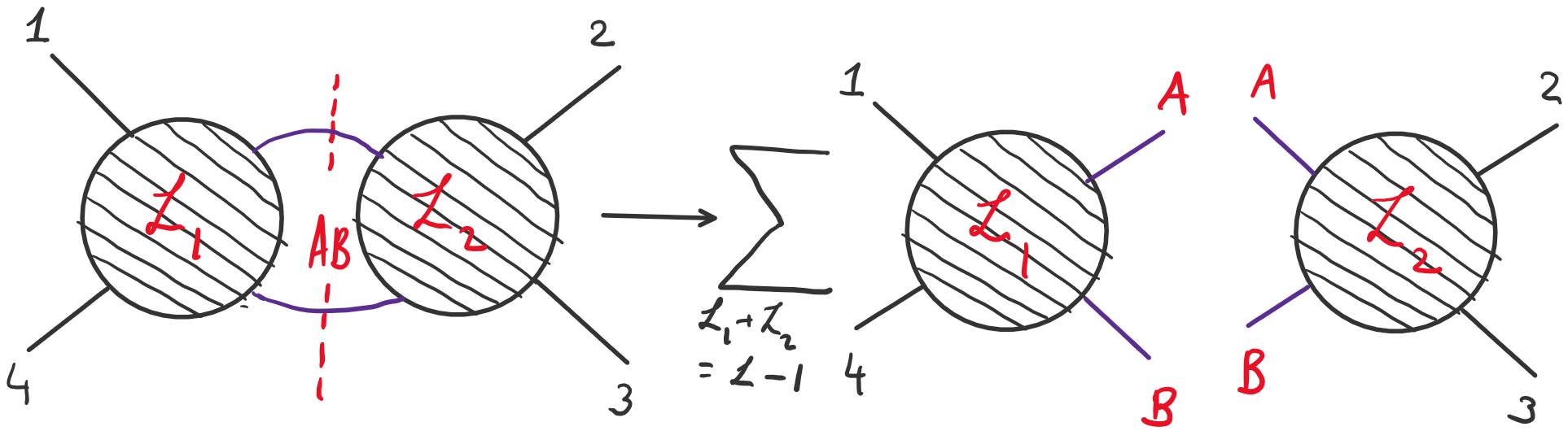}
\caption{Structure of the unitarity cut at 4 points}
\end{center}
\end{figure}
In this section we will examine the unitarity cut $\langle AB12 \rangle = \langle AB34 \rangle = 0$, at four points and show that the residue can be written as a product of lower loop, 4-point amplitudes. Specifically, we will show that the defining conditions of the amplituhedron (\ref{eq:kamplituhedrontree1}) - (\ref{eq:kamplituhedronloop2}) can be replaced by two disjoint set of conditions which define a ``left amplituhedron" with external data $\left\lbrace Z_1, A, B,Z_4\right\rbrace$ and a ``right" amplituhedron with external data $\left\lbrace A, Z_2, Z_3, B\right\rbrace$. We will show that the mutual positivity conditions in (\ref{eq:mutualpositivtycondition}) which seemingly connect the ``left" and ``right" amplituhedra are automatically satisfied once the defining conditions for the ``left" and ``right" amplituhedra are met. This suffices to prove that the canonical form on the cut is
\beas
\sum_{L_1+L_2 = L-1} \mathcal{M}_{\mathcal{L}}^{L_1}(Z_1, A, B, Z_4) \,\, \mathcal{M}_{\mathcal{R}}^{L_2}(B, A, Z_{2}, Z_3)
\eeas
where $\mathcal{M}_{\mathcal{L}}^{L_1}(Z_1, A, B, Z_4)$ and $\mathcal{M}_{\mathcal{R}}^{L_2}(B, A, Z_{2}, Z_3)$ are the canonical forms of the ``left" and ``right" amplituhedra respectively. A suitable parametrization of $(AB)$ is 
\bea
\label{eq:4ptcutpar}
A = Z_1 + x Z_2 \qquad B = y Z_3 +  Z_4
\eea
This ensures the cut conditions are satisfied. To compute the canonical form on the cut, we must solve the remaining inequalities. The tree level constraints (\ref{eq:kamplituhedrontree1}), (\ref{eq:kamplituhedrontree2}) trivially imply $\langle 1234\rangle >0$. The remaining loop level conditions in (\ref{eq:kamplituhedronloop1}) and (\ref{eq:kamplituhedronloop2}) impose $\langle AB13 \rangle <0$ and $\langle AB14 \rangle >0$ which ensure $x>0, y>0$. Denoting the uncut loops as $(AB)_a$ with $a=1, \dots L-1$, the remaining inequalities are loop positivity conditions,
\bea
\label{eq:leftoverloopconditiions}
\langle (AB)_a jj+1\rangle >\, 0 \qquad \forall j = 1, \dots 4,
\eea
mutual positivity among the uncut loops  
\bea
\label{eq:uncutloopmutualpos}
\langle (AB)_a (AB)_b \rangle > 0 ,
\eea
and mutual positivity with the cut loop
\bea
\label{eq:mutual positivity}
\langle ABA_aB_a \rangle = \langle A_a B_a 13\rangle y + \langle A_a B_a 14\rangle + \langle A_a B_a 23\rangle x y + \langle A_a B_a 24\rangle x >0 .
\eea
Here we have used the parametrization $(\ref{eq:4ptcutpar})$ for $A$ and $B$. The consequences of this inequality are best understood by considering the related quantity $\left( \langle (AB)_a2B \rangle \langle (AB)_aA3\rangle \right)$. Using $(\ref{eq:4ptcutpar})$ for $A$ and $B$, we can rewrite this as follows.
\bea
\label{eq:factorizationmutual}
&&\nonumber\left( \langle (AB)_a2B \rangle \langle (AB)_aA3\rangle \right)\\
&&\nonumber=\left( \langle A_aB_a24\rangle + \langle A_a B_a 23\rangle y \right) \left( \langle A_aB_a13\rangle + \langle A_a B_a 23\rangle x \right)\\
&&\nonumber = \langle ABA_aB_a \rangle \langle A_aB_a23\rangle -\langle A_aB_a 14\rangle \langle A_aB_a 23\rangle +\langle A_aB_a 13\rangle \langle A_aB_a 24\rangle  \\
&&\nonumber = \langle ABA_a B_a \rangle \langle A_aB_a23\rangle+ \langle (A_aB_a1\cap A_aB_a2) 34\rangle \\
&& = \langle ABA_a B_a \rangle \langle A_aB_a23\rangle+ \langle A_aB_a12 \rangle \langle A_aB_a 34\rangle 
\eea
The above equation implies $\langle (AB)_a2B \rangle \langle (AB)_aA3\rangle >0$ as each term on the right hand side is individually positive due to (\ref{eq:leftoverloopconditiions}) and (\ref{eq:mutual positivity}). The two possible solutions are
\bea
\label{eq:condition1}
\langle (AB)_a2B \rangle > 0 \qquad  \langle (AB)_aA3\rangle > 0
\eea
and
\bea
\label{eq:condition2}
\langle (AB)_a2B \rangle < 0 \qquad  \langle (AB)_aA3\rangle < 0
\eea
A particular loop $(AB)_a$ may satisfy either (\ref{eq:condition1}) or (\ref{eq:condition2}). In a generic case, there will be $L_1$ loops, $(AB)_{a_1}$ which obey $(\ref{eq:condition1})$ and $L_2 = L - L_1 - 1$ loops, $(AB)_{a_2}$ which obey $(\ref{eq:condition2})$. There are no restrictions on what values $L_1$ and $L_2$ can take. Consequently, the complete region satisfying the inequalities (\ref{eq:leftoverloopconditiions}) and (\ref{eq:mutual positivity}) is a sum over all values of $L_1$ and $L_2$ with $L_1+L_2=L-1$. We will now show that the canonical form for a region with fixed $L_1$ and $L_2$ can be written as a product of forms for amplituhedra $\mathcal{A}_{4,0,L_1}$ and $\mathcal{A}_{4,0,L_2}$. From Fig. \ref{fig:4pt}, it is clear the the external data corresponding to the left amplitude is the set $\left\lbrace Z_1,A,B,Z_4\right\rbrace$. Any loops $(AB)_{a_1}$ which belongs to this amplituhedron must satisfy the defining conditions \ref{eq:kamplituhedrontree1}, \ref{eq:kamplituhedrontree2}, \ref{eq:kamplituhedronloop1}, \ref{eq:kamplituhedronloop2}. 
\bea
\label{eq:L1conditions}
\text{\underline{Tree Level}} &&\langle 1AB4 \rangle = \langle AB14 \rangle = \langle 1234\rangle >0 \text{ from } \ref{eq:4ptcutpar}\\
\text{\underline{Loop level}} &&\nonumber \langle (AB)_{a_1} 1A \rangle =x \,\langle (AB)_{a_1} 12\rangle >0 \qquad \langle (AB)_{a_1} B4 \rangle =y \langle (AB)_{a_1} 34\rangle >0\\
&&\nonumber \text{ Both of these follow from (\ref{eq:4ptcutpar}) and (\ref{eq:leftoverloopconditiions}})\\
&&\nonumber \langle (AB)_{a_1} \, AB \rangle > 0 \text{ from } (\ref{eq:mutual positivity}) \\
&&\nonumber\text{The sequence} \left\lbrace \langle (AB)_{a_1} 1A \rangle, \langle (AB)_{a_1} 1B \rangle, \langle (AB)_{a_1} 14 \rangle \right\rbrace \text{has 2 sign flips}\\ 
\nonumber\text{\underline{Mutual positivity}} &&\langle (AB)_{a_1} (AB)_{a'_1} \rangle > 0 \,\,\, (\ref{eq:uncutloopmutualpos}), 
\eea
The flip condition is the only one left to be verified and follows from the Pl$\ddot{\text{u}}$cker relation
\bea
\label{eq:plucker}
\langle (AB)_{a_1} 1B \rangle \langle (AB)_{a_1}23 \rangle - \langle (AB)_{a_1} 2B \rangle \langle (AB)_{a_1}13 \rangle = \langle (AB)_{a_1} 12 \rangle \langle (AB)_{a_1} B3\rangle 
\eea
This can be derived by noting that the 5 twistors $\left\lbrace B, A_{a_1}, B_{a_1}, Z_1, Z_2 \right\rbrace$ are linearly dependent which leads to the condition
\beas
\langle BA_{a_1}B_{a_1}1\rangle Z_2 \,+\, \langle A_{a_1}B_{a_1}12\rangle B \, + \, \langle B_{a_1}12B\rangle A_{a_1} \,+ \,\langle 12BA_{a_1}\rangle B_{a_1} \, + \, \langle 2BA_{a_1}B_{a_1}\rangle Z_1 = 0.
\eeas 
Contracting this with $(AB){a_1}Z_3$ yields (\ref{eq:plucker}). Note that the RHS of $(\ref{eq:plucker})$ is negative since $\langle (AB)_{a_1} B3 \rangle = -\langle (AB)_{a_1} 34\rangle <0$ while the signs of the terms in the LHS are  
\bea
&&\left\lbrace \langle (AB)_{a_1} 23\rangle, \langle (AB)_{a_1}2B\rangle, \langle (AB)_{a_1}13\rangle \right\rbrace\\
&&\nonumber \left\lbrace \hspace{1cm}+\hspace{1cm}, +\hspace{1cm},\hspace{1cm} -\hspace{.85cm} \right\rbrace
\eea
This forces $\langle (AB)_a 1B \rangle <0$ and ensures that the sequence in (\ref{eq:L1conditions}) has 2 flips.\\

Similarly, the external data for the right is the set $\left\lbrace A, Z_2, Z_3, B \right\rbrace$ and a loop $(AB)_{a_2}$ which belongs to it satisfies the following conditions.
\bea
\label{eq:L2conditions}
\text{\underline{Tree Level}} &&\langle A23B \rangle = \langle AB23 \rangle >0\\
\text{\underline{Loop level}} &&\nonumber \langle (AB)_{a_2} A2 \rangle =\langle (AB)_{a_2} 12\rangle >0 \qquad \langle (AB)_{a_2} 23 \rangle > 0\\
&&\nonumber\langle (AB)_{a_2} 3B \rangle =\langle (AB)_{a_2} 34\rangle >0 \\
&&\nonumber\text{The sequence} \left\lbrace \langle (AB)_{a_2} A2 \rangle, \langle (AB)_{a_2} A3 \rangle, \langle (AB)_{a_2} AB \rangle \right\rbrace \text{has 2 sign flips}\\ 
\nonumber\text{\underline{Mutual positivity}} &&\langle (AB)_{a_2} (AB)_{a'_2} \rangle > 0, 
\eea
Clearly, the conditions ($\ref{eq:L1conditions}$) and ($\ref{eq:L2conditions}$) define the amplituhedra $\mathcal{A}_{4,0,L_1}$ and $\mathcal{A}_{4,0,L_2}$ with canonical forms $\mathcal{M}_{\mathcal{L}}^{L_1}(Z_1, A, B, Z_4)$ and $\mathcal{M}_{\mathcal{R}}^{L_2}(B, A, Z_{2}, Z_3)$ respectively. To complete the proof that the canonical form on the cut is just the product of these forms, we must show that that mutual positivity between the loops $(AB)_{a_1}$ and $(AB)_{a_2}$ imposes no further constraints. To see this, we can expand the loop $(AB)_{a_1}$ in terms of $\left\lbrace Z_1, A, B ,Z_4 \right\rbrace$ 
\beas
A_{a_1} = Z_1 + \alpha_1 A + \alpha_2 B  \qquad B_{a_1} = -Z_1 + \beta_1 B +\beta_2 Z_4
\eeas
and compute $\langle (AB)_{a_1} (AB)_{a_2} \rangle$ which yields, 
\bea
\nonumber
\langle (AB)_{a_1} (AB)_{a_2} \rangle =&& y \langle (AB)_{a_2} 1B \rangle \beta_1 + \langle (AB)_{a_2} 14 \rangle \beta_2 + \langle (AB)_{a_2} 1A \rangle (\alpha_1) + \langle (AB)_{a_2} AB \rangle \alpha_1 \beta_1 \\
\label{eq:mutualpos}
&&+ \langle (AB)_{a_2} A4 \rangle {a_2}_1 \beta_2 + \langle (AB)_{a_2} 1B \rangle ({a_2}_2) + \langle (AB)_{a_2} B4 \rangle {a_2}_2 \beta_2
\eea
The positivity of all the terms except for $\langle (AB)_{a_2} A4 \rangle $ and $\langle (AB)_{a_2} B1 \rangle$ immediately follows from (\ref{eq:L2conditions}). For these two, we have 
\beas
&&\langle (AB)_{a_2} A4 \rangle = \langle (AB)_{a_2} A(B-y 3) \rangle = \langle (AB)_{a_2} AB \rangle - y \langle (AB)_{a_2} A3\rangle >0 \\
&&\langle (AB)_{a_2} 1B \rangle = \langle (AB)_{a_2} (A-x 2)B \rangle = \langle (AB)_{a_2} AB \rangle - x \langle (AB)_{a_2} 2B\rangle >0  
\eeas
Therefore, $\langle (AB)_{a_1}(AB)_{a_2} \rangle > 0$ imposes no new constraints and the canonical form on the cut factorizes into $\mathcal{M}_{\mathcal{L}}$ and $\mathcal{M}_{\mathcal{R}}$.

\section{Proof for MHV amplitudes of arbitrary multiplicity} 
\label{sec:MHVproof}
We will extend the above results to amplitudes of arbitrary multiplicity. However, the existence of higher $k$ sectors beginning with $n=5$ complicates the proof. In this section we will focus on a proof of unitarity for MHV amplitudes. This allows us to sketch the essentials of the proof without additional complications. In the next section, we modify the proof to account for higher $k$ sectors.\\

We are examining the residue of the MHV amplituhedron $\mathcal{A}_{n,0,L}\left(Z_1, \dots, Z_n\right)$ on the cut $\langle ABii+1 \rangle =  \langle ABjj+1 \rangle = 0$. For the rest of the paper, we will assume $j\neq i+1$\footnote{In the singular case of $j=i+1$, the amplitude factorizes into a 3-point MHV or $\overline{\text{MHV}}$ amplitude a $n+1$ point MHV amplitude. The 3 point case is degenerate and the use of momentum twistors ensures that all the defining conditions(on-shell and momentum conservation) are always satisfied.  There are no further constraints that need to be imposed}. The defining conditions for the amplituhedron $\mathcal{A}_{ n,0,L}\lbrace Z_1, \dots Z_n \rbrace$ are
\bea
\label{eq:MHVnpointamplituhedron}
\text{\underline{Tree Level}} &&\langle ijkl\rangle >0 \text{ for } i<j<k<l\\
\nonumber\text{\underline{Loop level}} &&\langle ii+1 \rangle >0 \\
\nonumber &&\text{The sequence }  S=\left\lbrace \langle i+1i+2 \rangle, \dots \langle i+1n \rangle, -\langle i+11\rangle, \dots -\langle i+1i\rangle \right\rbrace\\
\nonumber && \text{has 2 sign flips. }\\
\nonumber\text{\underline{Mutual Positivity}} &&\langle (AB)_a(AB)_b \rangle > 0 \qquad \forall a, b \in \left\lbrace 1, \dots L\right\rbrace
\eea 
Here, $\langle ij \rangle \equiv \langle (AB)_a ij \rangle$. We have chosen to look at the flip pattern of a particularly convenient sequence. All other related sequences will also have the same number of flips as mentioned in Section [\ref{sec:looplevelconditions}].\\

There are clearly many patterns of signs for which the sequences $S$ has two sign flips. We refer to each pattern as a configuration of the amplituhedron. A configuration for the MHV amplituhedron is specified by giving the signs of all entries of the sequence S. We would like to show that for each configuration, the canonical form can be written as the product of canonical forms of a left and right amplituhedron. To begin, we we can parametrize the cut loop $AB$ as
\bea
\label{eq:cutpar}
A = Z_i + x \, Z_{i+1} \qquad B = y \, Z_j + Z_{j+1} \qquad 
\eea  
To show that the canonical form on this cut can written as a product of canonical forms for lower loop, ``left" and ``right" MHV amplituhedra $\mathcal{A}^{\mathcal{L}}_{n_1,0,L_1}$ and $\mathcal{A}^{\mathcal{R}}_{n_2,0,L_2}$ (with $L_2 = L - L_1 - 1$), we need precise definitions of these objects. This is provided in the following section. 
\subsection{Left and right Amplituhedra}
\subsubsection{The left amplituhedron $\mathcal{A}_{n_1,0,L_1}$}
\label{sec:left amp}
\begin{figure}[htb!]
\begin{center}
\label{fig:leftamp}
\includegraphics[scale=.5]{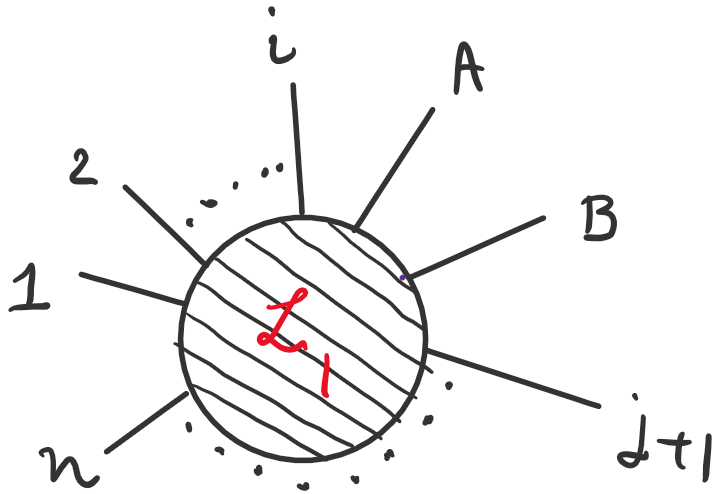}
\caption{The left amplituhedron}
\end{center}
\end{figure}
The left amplituhedron $\mathcal{A}_{n_1, 0, L_1}$ is defined by three sets of conditions similar to $(\ref{eq:MHVnpointamplituhedron})$. In this case, the external data, as seen from Fig \ref{fig:leftamp} is the set $\mathcal{L}=\left\lbrace Z_1, \dots Z_i, A, B, Z_{j+1}, \dots , Z_n\right\rbrace$. Letting $a,b, c, d$ denote elements of this set  and $\langle ij \rangle \equiv \langle (AB)_a ij\rangle$, the defining conditions are
\bea
\label{eq:leftconditions}
\text{\underline{Tree Level}}&&\nonumber  \forall  a<b<c<d \in \mathcal{L}\\
&&\nonumber \langle abcd \rangle >0,\,\, \, \langle iAab \rangle >0,\,\, \langle ABab \rangle >0, \, \, \langle Bj+1ab\rangle >0
\eea
These are satisfied if $x>0$ and $y>0$. 
\bea
\text{\underline{Loop Level}}\nonumber &&\langle aa+1\rangle >0, \, \langle iA \rangle >0,\, \langle AB \rangle >0,\, \langle Bj+1 \rangle >0\,\\
&&\nonumber \text{The sequence } S_L = \left\lbrace \langle iA\rangle, \langle iB\rangle, \langle ij+1\rangle, \dots  \langle in\rangle, -\langle i1\rangle, \dots -\langle ii-1\rangle \right\rbrace\\
&&\nonumber \text{has 2 sign flips}\\
\underline{\text{Mutual Positivity}}\nonumber&&\langle (AB)_a (AB)_b \rangle >0 .
\eea
The above sequence lends itself to easy comparison with the sequence $S$ in (\ref{eq:MHVnpointamplituhedron}). However, for consistency, we must also verify that the following sequences have the same number of sign flips as $S_L$. 
\begin{align*}
& \lbrace \langle AB\rangle, \langle Aj+1\rangle, \dots , -\langle Ai \rangle \rbrace\\
&\lbrace \langle Bj+1 \rangle, \langle Bj+2\rangle,  \dots , -\langle BA \rangle  \rbrace \\
&\qquad\vdots\\
&\lbrace \langle i-1i\rangle, \langle i-1A\rangle, \dots -\langle i-1i-2\rangle\rbrace 
\end{align*}
This ensures that the definition of the amplituhedron is independent of the choice of sequence, similar to Section [\ref{sec:looplevelconditions}]. The positivity conditions on the loop data ensures that all the first and last entries of these sequences are positive. Furthermore any two sequences in the above set, all of which are of the form $\left\lbrace \langle ak \rangle \right\rbrace$ and $\left\lbrace \langle a+1k \rangle \right\rbrace$, satisfy
\bea
\label{eq:pluckernpoint}
\langle ak\rangle \langle a+1k+1\rangle - \langle ak+1\rangle \langle a+1k\rangle = \langle aa+1\rangle \langle kk+1\rangle >0  
\eea 
The equality of sign flips now follows from the analysis in Appendix $[\ref{sec:app1}]$. This shows that the left amplituhedron can be consistently defined at tree level. The mutual positivity and the loop level positivity conditions for all the loops in the left amplituhedron are automatically satisfied because of $(\ref{eq:cutpar})$ and $(\ref{eq:MHVnpointamplituhedron})$. The flip condition defines the criterion for any uncut loop $(AB)_a$ to be in the left amplituhedron. We will present a detailed analysis in Section [\ref{sec:MHVfactorization}].\\
\subsubsection{The right Amplituhedron $A_{n_2,0,L_2}$}
\begin{figure}[htb!]
\begin{center}
\includegraphics[scale=.5]{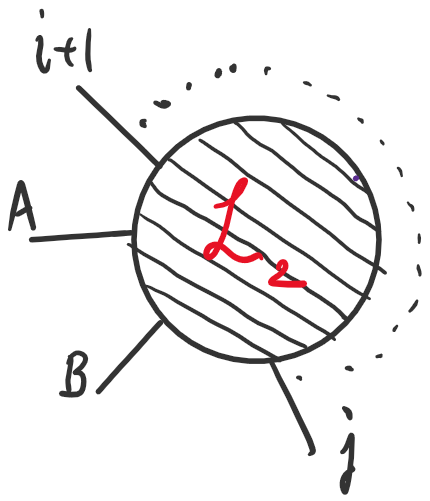}
\caption{The right amplituhedron}
\end{center}
\end{figure}

The external data for the right amplituhedron $\mathcal{A}_{n_2, 0, L_2}$ is $\mathcal{R} = \lbrace A, Z_{i+1},\dots , Z_j, B\rbrace$ and the defining inequalities are listed below. $a, b, c, d \in \mathcal{R}$ and $\langle ij\rangle \equiv \langle (AB)_aij\rangle$ with $(AB)_a$ being an uncut loop.
\bea
\label{eq:rightconditions}
\text{\underline{Tree Level}}&&\langle abcd\rangle > 0,\,\,\, \langle Ai+1ab\rangle >0, \,\,\, \langle abjB \rangle >0, \\
&& \nonumber\langle ABab\rangle >0 \qquad \text{ with }a<b<c<d\\
\nonumber\text{\underline{Loop Level}}&&\langle Ai+1 \rangle >0,\,\,\, \langle jB \rangle > 0, \,\,\,\langle aa+1 \rangle >0\\
&&\nonumber\text{The sequence }S_R = \left\lbrace \langle i+1i+2\rangle, \dots \langle i+1j\rangle, \langle i+1B\rangle, -\langle i+1A \rangle \right\rbrace \\
&&\nonumber\text{has 2 sign flips}\\
\text{\underline{Mutual Positivity}}&&\nonumber\langle (AB)_a (AB)_b \rangle >0
\eea
Once again, for consistency we should verify that
\begin{align*}
&S_1: \lbrace \langle Ai+1\rangle, \langle Ai+2\rangle, \dots , \langle AB \rangle \rbrace\\
&S_2: \lbrace \langle i+2i+3 \rangle, \langle i+2i+4\rangle,  \dots , -\langle i+2i+1 \rangle  \rbrace \\
\vdots\\
&S_{2+j-i}: \lbrace -\langle BA\rangle, -\langle Bi+1\rangle, \dots -\langle Bj\rangle\rbrace 
\end{align*}
all have the same number of sign flips as $S_R$. The proof is identical to the one for the left amplituhedron. The tree level, mutual positivity and loop level positivity conditions are once again guaranteed by  $(\ref{eq:cutpar})$ and $(\ref{eq:MHVnpointamplituhedron})$ and an analysis of the flip condition is in Section [\ref{sec:MHVfactorization}].

\subsection{Factorization on the unitarity cut}
\label{sec:MHVfactorization}
It was shown in the last section that the two sets $\mathcal{L} = \left\lbrace Z_1, \dots, Z_i, A, B, Z_{j+1}, \dots, Z_n\right\rbrace$ and $\mathcal{R} = \left\lbrace A, Z_{i+1}, \dots Z_j, B\right\rbrace$ define positive external data and that the loop level positivity conditions are satisfied. We need to analyze every configuration of the amplituhedron and show that for each configuration, an uncut loop belongs to the left or the right amplituhedron. The similar analysis for the 4 point case, performed in Section [\ref{sec:4pt}], was much simpler owing to the fact there there was only one possible sign pattern for the sequence, $\left\lbrace \langle AB12\rangle, \langle AB13\rangle, \langle AB14\rangle \right\rbrace$. Here, the presence of multiple compatible sign patterns increases the complexity of the proof and no simple relation like (\ref{eq:factorizationmutual}) exists. It is natural to label the configurations of $\mathcal{A}_{n,0,L}(Z_1, \dots Z_n)$ by looking at the sign patterns in the sequence S as explained below.
\beas
S = \left\lbrace \langle i+1i+2 \rangle,\quad \dots \quad \langle i+1j \rangle \,\vline\, \langle i+1j+1\rangle, \quad\dots\quad -\langle i+1i\rangle \right\rbrace
\eeas
Note that the flip pattern of this sequence determines whether the loop $(AB)_a$ belongs to the original amplituhedron which has external data $\left\lbrace Z_1, \dots Z_n\right\rbrace$. Consequently, it doesn't involve the points $A$ and $B$. We have divided the sequence in a suggestive way. The ${\it left}$ half of $S$ looks very similar to $S_R$ (\ref{eq:rightconditions}). It is natural to label the different flip patterns of $S$ as $S_{ablr}$ where $a,b = \pm$ are the signs of $\langle i+1j\rangle $ and $\langle i+1j+1\rangle$ and $l,r$ are the number of flips in the left and right parts of $S$. \\

In order to compare $S_L$ (\ref{eq:leftconditions}) to $S$, we introduce the sequence 
\bea
\label{eq:SL'compare}
S'_L  = \left\lbrace \langle i+1A\rangle, \langle i+1B\rangle, \langle i+1j+1\rangle, \dots -\langle i+1i-1\rangle \right\rbrace
\eea 
and call the number of flips in this sequence $k'_L$ flips. The motivation behind introducing this is that $S_L$ and $S'_L$ are connected by the Pl$\ddot{\text{u}}$cker relation (similar to (\ref{eq:pluckernpoint}))
\bea
\langle ik\rangle \langle i+1k+1\rangle - \langle ik+1\rangle \langle i+1k\rangle = \langle ii+1\rangle \langle kk+1\rangle \, >\,  0
\eea 
Following the analysis in Appendix[$\ref{sec:app1}$], the relation between $k_L$ and $ k'_L$ is determined entirely by the signs of the first and last elements
\beas
\begin{pmatrix}
\langle iA\rangle & -\langle ii-1\rangle\\
\langle i+1A \rangle & -\langle i+1i-1\rangle
\end{pmatrix} = \begin{pmatrix}
+ &  +\\
- & \langle i-1i+1\rangle
\end{pmatrix} 
\eeas
where $k_L$ is the number of sign flips in $S_L$. If $\langle i-1i+1\rangle >0$, then $k_L = k'_L-1$ otherwise $k_L = k'_L$.\\

$S$ now looks almost like a juxtaposition of $S_R$ and $S'_L$. Each flip pattern of S determines whether the corresponding loop $(AB)_a$ belongs to the left or the right amplituhedron as shown below.

\begin{itemize}
\item $S_{++20} = \left\lbrace +, \dots \text{2 flips} \dots +\vline +\dots \text{0 flips} \dots + \right\rbrace$
\end{itemize}
The sequence $S_R$ clearly has 2 sign flips since 
\beas
-\langle i+1A\rangle > 0 \text{ and } \langle i+1j\rangle >0, \, \langle i+1j+1\rangle >0 \implies \langle i+1B \rangle >0
\eeas 
$S'_L$ has one sign flip since 
\beas
\langle i+1A \rangle >0,\, \langle i+1B\rangle >0, \langle i-1i+1\rangle >0.
\eeas 
Furthermore $k_L = k'_L-1 = 0$ and the loop $(AB)_a$ belongs only to the right amplituhedron. 

\begin{itemize}
\item $S_{++02} = \left\lbrace +, \dots \text{0 flips }\dots +\vline + \dots \text{2 flips }\dots +\right\rbrace$
\end{itemize}
$S_R$ obviously has $0$ sign flips. If $\langle i-1i+1 \rangle > 0$, $k_L = k'_L+1 = 2$ and if $\langle i-1i+1\rangle < 0$, $k'_L=k_L=2$. In both cases, the loop belongs to the left amplituhedron and not the right. 

\begin{itemize}
\item $S_{+-01} = \left\lbrace +, \dots \text{0 flips} \dots +\vline - \dots\text{1 flip} \dots +\right\rbrace$
\end{itemize}
If $\langle i+1B \rangle >0$, then the sequence $S_R$ has 0 flips and the loop doesn't belong to the right amplituhedron. If $\langle i-1i+1\rangle >0$, $k'_L=3$ and $k_L = k'_L-1 = 2$. Otherwise, $k'_L = 2$ and $k_L = k'_L = 2$. Thus irrespective of the sign of $\langle i-1i+1\rangle$, the loop $(AB)_a$ belongs to the left amplituhedron. \\

If $\langle i+1B \rangle <0$, then $S_R$ has 2 sign flips and it can be shown that $k_L=0$ by analysis similar to the cases above. This $(AB)_a$ belongs to the right amplituhedron. 

\begin{itemize}
\item $S_{-+10} = \left\lbrace +, \dots \text{1 flip }\dots -\vline + \dots \text{0 flips }\dots +\right\rbrace$
\end{itemize}
In this case, $S_R$ has two flips and $S'_L$  has one flip irrespective of the sign of $\langle i+1B \rangle$. Since $k_L=k'_L-1$, we have $k_L=0$ and the loop belongs to the right amplituhedron.

\begin{itemize}
\item $S_{--11} = \left\lbrace +, \dots \text{1 flip }\dots -\vline - \dots \text{1 flip }\dots +\right\rbrace$
\end{itemize}
Once again, it is simple to show that $S_R$ has two sign flips and $S_L$ has 0 sign flips in this configuration.

\subsubsection{Trivialized mutual positivity}
We have shown that for every configuration of the amplituhedron, each loop belongs either to the left or the right. While we can consistently define left and right amplituhedra, it remains to be shown that the mutual positivity between a loop $(AB)_\mathcal{L}, \,(\mathcal{L} = 1, \dots L_1)$ in the left amplituhedron and a loop $(AB)_\mathcal{R}\,\, (\mathcal{R} = 1, \dots L_2)$ in the right amplituhedron doesn't impose any extra constraints. \\

This is easiest to see if we expand each loop $(AB)_\mathcal{L}$ and $(AB)_\mathcal{R}$ using (\ref{eq:Kermit}) as  
\beas
&& A_\mathcal{R} = A + \alpha_1 Z_{r_1} + \alpha_2 Z_{r_1+1} \qquad B_\mathcal{R} = -A + \beta_1 Z_{r_2} + \beta_2 Z_{r_2+1}\\
&& A_\mathcal{L} = A + \alpha_3 Z_{l_1} + \alpha_4 Z_{l_1+1} \qquad B_\mathcal{L} = -A + \beta_3 Z_{l_2} + \beta_4 Z_{l_2+1}
\eeas
with $r_1<r_2 \in \left\lbrace A, Z_{i+1}, \dots , Z_j, B \right\rbrace$ and $l_1<l_2 \in \left\lbrace Z_1, \dots Z_i, A, B, Z_{j+1}, \dots, Z_n \right\rbrace$. On expanding $\langle (AB)_\mathcal{L}(AB)_\mathcal{R}\rangle$, every term is of the form $\langle l_1l_2r_1r_2 \rangle$ with $\l_1<l_2<r_1<r_2$. Since the external data are positive, i.e. $\langle ijkl \rangle >0$ for $i<j<k<l$, we are assured that $\langle (AB)_\mathcal{L} (AB)_\mathcal{R}\rangle >0$.\\

\noindent This completes the proof of factorization for MHV amplituhedra on the unitarity cut. In the next section, we will extended this proof to the higher $k$ sectors.

\section{Proof for higher $k$ sectors}
\label{sec:higherk}
The proof of unitarity for higher $k$ is similar in spirit to that for the MHV sector. However, there are a lot additional details that we must take into account. Firstly, we must modify (\ref{eq:Unitarity}) to include products of ``left" and ``right" amplituhedra with different $k$. Suppose the left amplitude has $g_L$ negative helicity gluons and the right amplitude has $g_R$ negative helicity gluons, then we have $g_L + g_R = g+2$. With the MHV degrees defined as $k_L = g_L-2, k_R = g_R-2, k = g-2$, this equation reads $k_L+k_R = k$. Recall that we introduced the function $f(x,y,0,0, (AB)_a)$ in (\ref{eq:cut}) and stated the optical theorem in terms of it. Including sectors of different $k$, this becomes,
\bea
\label{eq:NKMHVoptical}
f(x,y,0,0,(AB)_a) =\sum_{k_L+k_R=k} \sum_{L_1+L_2 = L-1} \mathcal{M}_{\mathcal{L}}^{k_L, L_1} \,\mathcal{M}_{\mathcal{R}}^{k_R, L_2}
\eea
\begin{figure}[htb!]
\label{fig:NKMHV}
\includegraphics[scale=.4]{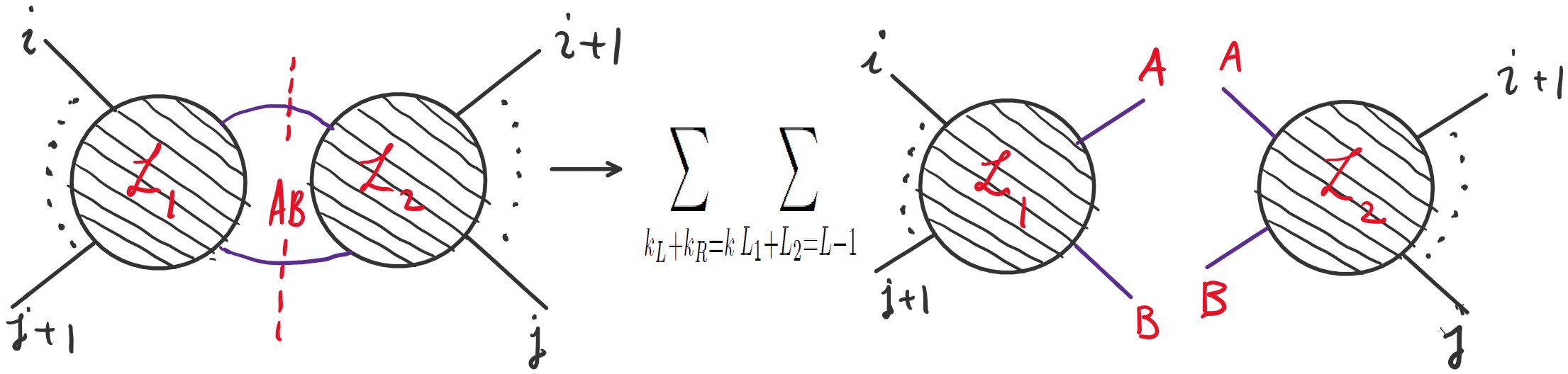}
\caption{Unitarity cut for an N$^k$MHV amplitude}
\end{figure}\\
We expect that unitarity emerges from a factorization property of the geometry in a manner similar to the MHV case. In order to make this statement more precise, we will have to define analogues of the left and right MHV amplituhedra for N$^k$MHV external data. $\mathcal{A}_{n,k,L}$, the $N^k$MHV amplituhedron defined by the conditions 
\bea
\label{eq:kamplituhedron}
\text{\underline{Tree level}}&&	\langle ii+1jj+1 \rangle > 0, \,\,\langle ii+1n1\rangle (-1)^{k-1}>0\text{ and the sequence }\\ 
\nonumber &&S^{\text{tree}}:\left\lbrace \langle ii+1i+2i+3 \rangle, \dots \langle ii+1i+2i-1\rangle (-1)^{k-1} \right\rbrace \text{ has $k$ sign flips.}\\
\text{\underline{Loop level}}&&\nonumber \langle (AB)_aii+1 \rangle >0, \,\, \langle ABn1\rangle (-1)^{k-1}>0 \text{ and the sequence }\\ 
&&\nonumber S^{\text{loop}}: \left\lbrace \langle (AB)_a 12\rangle, \langle (AB)_a 13\rangle, \dots \langle (AB)_a 1n\rangle \right\rbrace
\text{ has $k+2$ sign flips.}\\
\text{\underline{Mutual Positivity}}&&\nonumber \langle (AB)_a(AB)_b \rangle >0 
\eea
We can use any sequence $\left\lbrace \langle ABki\rangle \right\rbrace$ instead of $\left\lbrace \langle AB1i\rangle \right\rbrace$ as explained in Section [\ref{sec:looplevelconditions}].\\ 

It is worth re-emphasizing that we wish to prove that the canonical form on the cut (which is computed by solving the inequalities (\ref{eq:kamplituhedron}) for the uncut loops $(AB)_a$ along with 
$\langle ABii+1\rangle = \langle ABjj+1\rangle =0$) can be written as in (\ref{eq:NKMHVoptical}). For this to happen, we want to show that the set of inequalities in (\ref{eq:kamplituhedron}) can be replaced by two sets of inequalities which define lower loop amplituhedra, $\mathcal{A}_{n_1,k_L,L_1}^{\mathcal{L}}$ and $\mathcal{A}_{n_2,k_R,L_2}^{\mathcal{R}}$. It is not essential that the external data for these is a subset of $\left\lbrace Z_1, \dots Z_n \right\rbrace$. In particular they can be rescaled by factors $Z_i \rightarrow \sigma(i) Z_i$ and still yield the same canonical form due to projective invariance as discussed in Section [\ref{sec:amps}]. In fact, as we show below, this rescaling plays a crucial role in ensuring that the left and right amplituhedra have $k$ of both even and odd parity. \\

On the unitarity cut ($\langle ABii+1 \rangle = \langle ABjj+1\rangle = 0$ with $i+1\neq j$), there is a natural division of the external data into ``left" and ``right" sets, $\left\lbrace Z_1, \dots, Z_i, A, B, Z_{j+1}, \dots, Z_n \right\rbrace$ and $\left\lbrace A, Z_{i+1}, \dots, Z_j, B \right\rbrace$. However, insisting that this be the external data for the left and right amplituhedra imposes too many constraints. To see this, suppose that the ``left" set has MHV degree $k_L$. We must have $\langle ABn1\rangle (-1)^{k_L-1}>0$. But (\ref{eq:kamplituhedron}) implies that $\langle ABn1\rangle (-1)^{k-1}>0$. This forces $(-1)^{k+k_L}>0$ and restricts $k_L$ to be the same parity as $k$. Similarly for the right set, we have $\langle j-1jBA\rangle (-1)^{k_R-1}>0$ and again (\ref{eq:kamplituhedron}) implies $\langle ABj-1j\rangle >0$ which forces $(-1)^{k_R}>0$. In order to avoid these extra constraints on $k_L$ and $k_R$, we must allow for arbitrary signs on the $Zs$ and define two the sets of external data as
\bea
\label{eq:leftrightdata}
&&\nonumber \mathcal{L} = \left\lbrace\sigma_L(1) Z_1, \dots \sigma_L(i) Z_i, \sigma_L(A) A,\sigma_L(B) B,\sigma_L(j+1) Z_{j+1}, \dots ,\sigma_L(n) Z_n \right\rbrace\\
&&\mathcal{R}= \left\lbrace \sigma_R(A) A, \sigma_R(i+1) i+1, \dots, \sigma_R(j) j,\sigma_R(B) B \right\rbrace
\eea
where $\sigma(k) = \pm 1$. These signs will be determined by conditions like $(\ref{eq:kamplituhedron})$ which define the left and right amplituhedra along with the appropriate twisted cyclic symmetry. We will then show that the canonical form for every configuration in $\mathcal{A}_{n,k,L}$ can be mapped into a product of canonical forms on suitably defined left and right amplituhedra $\mathcal{A}^{\mathcal{L}}_{n_1,k_L,L_1}$ and $\mathcal{A}^{\mathcal{R}}_{n_2,k_R,L_2}$.
\subsection{The left and right amplituhedra} 
\label{sec:leftrightamp}
\subsubsection{The left amplituhedron $\mathcal{A}^{\mathcal{L}}_{n_1,k_L,L_1}$}
We must demand that the set $\mathcal{L}$ satisfies all the conditions in $(\ref{eq:kamplituhedron})$. In addition, this must also be compatible with the fact that the $Z_i$ are the external data for $\mathcal{A}_{n,k,L}$.
\bea
\label{eq:klefconditions}
&&\nonumber\langle aa+1bb+1\rangle \sigma_L(a)\sigma_L(a+1)\sigma_L(b)\sigma_L(b+1)>0\\
&&\langle ABaa+1\rangle \sigma_L(A)\sigma_L(B)\sigma_L(a)\sigma_L(a+1) >0\\
&&\nonumber \forall \,a,b \in \lbrace 1, \dots, i-1, j+1, \dots, n-1\rbrace
\eea
$AB$ and the $Zs$ automatically satisfy $\langle aa+1bb+1\rangle >0$ and $\langle ABaa+1\rangle >0$. Thus we have, $\sigma_L(a)\sigma_L(a+1)\sigma_L(b)\sigma_L(b+1) > 0$ and $\sigma_L(A)\sigma_L(B)\sigma_L(a)\sigma_L(a+1)>0$. 
Furthermore, we have new  constraints on $A$ and $B$ coming from 
\bea
\label{eq:leftregionconstraints}
&&\nonumber\langle iABj+1\rangle \sigma_L(i)\sigma_L(A)\sigma_L(B)\sigma_L(j+1) >0\\
&&\langle iAkk+1\rangle \sigma_L(i)\sigma_L(A)\sigma_L(k)\sigma_L(k+1) >0\\
&&\nonumber\langle Bj+1kk+1\rangle \sigma_L(B)\sigma_L(j+1)\sigma_L(k)\sigma_L(k+1) >0
\eea
Finally, since the set $\mathcal{L}$ is the external data for $\mathcal{A}^{\mathcal{L}}_{n_1,k_L,L_1}$, it must satisfy a twisted cyclic symmetry
\bea
\label{eq:lefttwisted}
&&\langle aa+1n1\rangle \sigma_L(a)\sigma_L(a+1)\sigma_L(n)\sigma_L(1)(-1)^{k_L-1} >0
\eea
Since $\langle aa+1n1\rangle (-1)^{k-1} >0$, consistency requires $(-1)^{k+k_L}\sigma_L(a)\sigma_L(a+1)\sigma_L(n)\sigma_L(1)>0$. This divides into two cases
\\
\begin{itemize}
\item $(-1)^{k+k_L}<0$
\end{itemize}
An allowed set $\lbrace\sigma_L(k)\rbrace$ satisfying $(\ref{eq:klefconditions})$ and $(\ref{eq:lefttwisted})$ is
\beas
&&\left\lbrace   \sigma_L(1), \dots , \sigma_L(i), \sigma_L(A), \sigma_L(B), \sigma_L(j+1), \dots, \sigma_L(n) \right\rbrace \\
&& = \left\lbrace   +, \hspace{.5cm}\dots\,\,\,,\quad  +,\hspace{.6cm} ?\hspace{.5cm},\hspace{.3cm} ?\hspace{.4cm},\hspace{.5cm} -\hspace{.7cm}, \dots ,\hspace{.5cm} -\hspace{.5cm}\right\rbrace 
\eeas
with $\sigma_L(A)$ and $\sigma_L(B)$ undetermined. $(\ref{eq:klefconditions})$ requires $\sigma_L(A)\sigma_L(B)>0$ and the constraints in $(\ref{eq:leftregionconstraints})$ read
\beas
\langle iABj+1\rangle <0 \qquad \langle iAkk+1\rangle \sigma_L(A)>0 \qquad \langle Bj+1kk+1\rangle \sigma_L(B)<0
\eeas 
The solutions to these constraints are
\beas
&&\mathcal{L}_1:\sigma_L(A) >0, \, \,\sigma_L(B)>0\, \, \text{ with } \,\langle iAkk+1 \rangle >0, \, \langle Bj+1kk+1\rangle <0, \, \langle iABj+1\rangle <0\\
&&\mathcal{L}_2:\sigma_L(A) <0, \, \,\sigma_L(B)<0\, \, \text{ with } \,\langle iAkk+1 \rangle <0, \, \langle Bj+1kk+1\rangle >0, \, \langle iABj+1\rangle<0
\eeas
\\
\begin{itemize}
\item $(-1)^{k+k_L}>0$
\end{itemize}
In this case $\lbrace\sigma_L(k)\rbrace$ satisfying $(\ref{eq:klefconditions})$ and $(\ref{eq:lefttwisted})$ is
\beas
&&\left\lbrace   \sigma_L(1), \dots , \sigma_L(i), \sigma_L(A), \sigma_L(B), \sigma_L(j+1), \dots, \sigma_L(n) \right\rbrace \\
&& = \left\lbrace   +, \hspace{.5cm}\dots\,\,\,,\quad  +,\hspace{.6cm} ?\hspace{.5cm},\hspace{.3cm} ?\hspace{.4cm},\hspace{.5cm} +\hspace{.7cm}, \dots ,\hspace{.5cm} +\hspace{.5cm}\right\rbrace 
\eeas
which again requires $\sigma_L(A)\sigma_L(B)>0$ and turns $(\ref{eq:leftregionconstraints})$ into
\beas
\langle iABj+1\rangle >0 \qquad \langle iAkk+1\rangle \sigma_L(A)>0 \qquad \langle Bj+1kk+1\rangle \sigma_L(B)>0.
\eeas 
This has the following solutions
\beas
&&\mathcal{L}_3:\sigma_L(A) >0, \, \,\sigma_L(B)>0\, \, \text{ with } \,\langle iAkk+1 \rangle >0, \, \langle Bj+1kk+1\rangle >0, \, \langle iABj+1\rangle >0\\
&&\mathcal{L}_4:\sigma_L(A) <0, \, \,\sigma_L(B)<0\, \, \text{ with } \,\langle iAkk+1 \rangle <0, \, \langle Bj+1kk+1\rangle <0, \, \langle iABj+1\rangle > 0
\eeas
Each of these regions is characterized by particular signs for $\langle iAkk+1\rangle$ and $\langle Bj+1kk+1\rangle$ along with a pattern of sign flips for the sequence 
\beas
S^{\text{tree}}_L: \left\lbrace \langle i-1iAB\rangle \sigma_L(B), \langle i-1iAj+1\rangle \sigma_L(j+1), \dots  \langle i-1iAi-2\rangle (-1)^{k_L-1}\sigma_L(i-2) \right\rbrace.
\eeas
Each region allows parametrization of the line $(AB)$ as $A = \pm Z_i \pm x Z_{i+1}$ and $B = \pm y Z_j \pm Z_{j+1}$ with $x>0, y>0$. In the table below, we list the different possibilities. \\

\begin{table}[htb!]
\label{table:leftregions}
\begin{center}
\begin{tabular}{|c|c|c|c|}
\hline
Region & $A$ & $B$ & $S_L^{\text{tree}}$ \\
\hline
$\mathcal{L}_1$ & $\pm Z_i + x Z_{i+1}$ & $-y Z_j \pm Z_{j+1}$ & $\left\lbrace +, \dots, (-1)^{k_L}\right\rbrace$\\
\hline
$\mathcal{L}_2$& $\pm Z_i - x Z_{i+1}$ & $y Z_j \pm Z_{j+1}$ & $\left\lbrace -, \dots, (-1)^{k_L-1}\right\rbrace$\\
\hline
$\mathcal{L}_3$& $\pm Z_i + x Z_{i+1}$ & $y Z_j \pm Z_{j+1}$ & $\left\lbrace +, \dots, (-1)^{k_L}\right\rbrace$\\
\hline
$\mathcal{L}_4$& $\pm Z_i - x Z_{i+1}$ & $-y Z_j \pm Z_{j+1}$ & $\left\lbrace -, \dots, (-1)^{k_L-1}\right\rbrace$\\
\hline
\end{tabular} 
\caption{Parametrization of $(AB)$ in the four regions}
\end{center}
\end{table}
It is crucial to remember that the canonical form is independent of the choice of $\sigma(i)$ and parametrization of $A$ and $B$. In all these cases the canonical form is that of $\mathcal{A}_{n_1, k_L, L_1}$.
\subsubsection{The right amplituhedron $\mathcal{A}^{\mathcal{R}}_{n_2,k_R,L_2}$}
A similar analysis of the effects of $(\ref{eq:kamplituhedron})$ on the set $\mathcal{R}$ yields the following constraints on $\lbrace\sigma_R\rbrace$.
\bea
\label{eq:krightconstraints}
&&\sigma_R(a)\sigma_R(a+1)\sigma_R(b)\sigma_R(b+1) >0 \\
&&\nonumber \sigma_R(A)\sigma_R(i+1)\sigma_R(k)\sigma_R(k+1)\langle Ai+1kk+1\rangle >0\\
&&\nonumber \sigma_R(j)\sigma_R(B)\sigma_R(k)\sigma_R(k+1)\langle jBkk+1\rangle >0\\
&&\nonumber  \sigma_R(B)\sigma_R(A)\sigma_R(k)\sigma_R(k+1) \langle BAkk+1\rangle (-1)^{k_L-1} >0\\
&&\nonumber \sigma_R(A)\sigma_R(B)\sigma(i+1)\sigma_R(j)\langle ABi+1j\rangle >0
\eea
These conditions are satisfied by 
\beas
&&\left\lbrace   \sigma_R(1), \dots , \sigma_R(i), \sigma_R(A), \sigma_R(B), \sigma_R(j+1), \dots, \sigma_R(n) \right\rbrace \\
&& = \left\lbrace   +, \hspace{.5cm}\dots\,\,\,,\quad  +,\hspace{.6cm} ?\hspace{.5cm},\hspace{.3cm} ?\hspace{.4cm},\hspace{.5cm} +\hspace{.7cm}, \dots ,\hspace{.5cm} +\hspace{.5cm}\right\rbrace 
\eeas
with $\sigma_R(A)$ and $\sigma_R(B)$ having solutions depending on $k_R$.
\begin{itemize}
\item $(-1)^{k_R}>0$
\end{itemize}
\beas
&&\mathcal{R}_1:\sigma_R(A) >0,\, \sigma_R(B)>0 \text{ with } \langle Ai+1kk+1\rangle>0, \,\, \langle jBkk+1\rangle >0, \, \langle ABi+1j \rangle >0\\
&&\mathcal{R}_2:\sigma_R(A) <0,\, \sigma_R(B)<0 \text{ with } \langle Ai+1kk+1\rangle<0, \,\, \langle jBkk+1\rangle <0, \, \langle ABi+1j \rangle >0
\eeas
\begin{itemize}
\item $(-1)^{k_R}<0$
\end{itemize}
\beas
&&\mathcal{R}_3:\sigma_R(A) >0,\, \sigma_R(B)<0 \text{ with } \langle Ai+1kk+1\rangle>0, \,\, \langle jBkk+1\rangle <0, \, \langle ABi+1j \rangle <0\\
&&\mathcal{R}_4:\sigma_R(A) <0,\, \sigma_R(B)>0 \text{ with } \langle Ai+1kk+1\rangle<0, \,\, \langle jBkk+1\rangle >0, \, \langle ABi+1j \rangle <0
\eeas
Once again, each region is characterized by different pattern of sign flips of the sequence 
\beas
S^{\text{tree}}_R:\left\lbrace \langle Ai+1i+2i+3 \rangle \sigma_R(i+3), \dots, \langle Ai+1i+2B\rangle \sigma_R(B) \right\rbrace 
\eeas
where we have ignored an overall factor of $\sigma_R(A)\sigma_R(i+1)\sigma_R(i+2)$. We list the various parametrizations and sign patterns of $S_R^{\text{tree}}$ below.\\ 
\begin{table}[htb!]
\label{table:rightregions}
\begin{center}
\begin{tabular}{|c|c|c|c|}
\hline
Region & $A$ & $B$ & $S_R^{\text{tree}}$ \\
\hline
$\mathcal{R}_1$ & $ Z_i \pm x Z_{i+1}$ & $\pm y Z_j + Z_{j+1}$ & $\left\lbrace +, \dots, +\right\rbrace$\\
$\mathcal{R}_2$& $ -Z_i \pm x Z_{i+1}$ & $\pm y Z_j - Z_{j+1}$ & $\left\lbrace +, \dots, +\right\rbrace$\\
$\mathcal{R}_3$& $ Z_i \pm x Z_{i+1}$ & $\pm y Z_j - Z_{j+1}$ & $\left\lbrace +, \dots, -\right\rbrace$\\
$\mathcal{R}_4$& $- Z_i \pm x Z_{i+1}$ & $\pm y Z_j + Z_{j+1}$ & $\left\lbrace -, \dots, +\right\rbrace$\\
\hline
\end{tabular} 
\caption{Parametrization of $(AB)$ in the four regions}
\end{center}
\end{table}
\\The canonical form is independent of the choice of $\sigma(i)$ and parametrization of $A$ and $B$.
\subsection{Factorization of the external data}
We will show that, on the unitarity cut, for every allowed sign flip pattern of the sequence $S^{\text{tree}}$, there exist regions $\mathcal{L}_i, \mathcal{R}_i$ such that $S^{\text{tree}}_L$ and $S^{\text{tree}}_R$ have the flip patterns necessary for $\mathcal{A}_{n_1,k_L,L_1}^{\mathcal{L}}$ and $\mathcal{A}_{n_2,k_R,L_2}^{\mathcal{R}}$. The analysis that follows is similar to the one is Section [\ref{sec:MHVfactorization}]. The sequence $S_R$ is similar to the {\it left} part of $S^{\text{tree}}$ and can be compared directly. In order to compare $S^\text{tree}_L$  with $S^\text{tree}$, it is necessary to introduce another sequence $S'^{\text{tree}}_L$. This is analogous to what we did in (\ref{eq:SL'compare}).
\beas
S^{'\text{tree}}_L : \left\lbrace \langle i+2iAB \rangle \sigma_L(B), \langle i+2iAj+1\rangle \sigma_L(j+1), \dots, \langle i+2iAi-2\rangle \sigma_L(i-2) (-1)^{k_L-1}\right\rbrace
\eeas
Let $k, k_L, k'_L, k_R$ be the number of flips in $S_L^{\text{tree}}, S_L^{'\text{tree}}, S_R^{\text{tree}}, S^{\text{tree}}$ respectively. $k_L$ and $k'_L$ are related to each other due to the following Pl$\ddot{u}$cker relations
\beas
&&\sigma_L(B)\sigma_L(j+1)\left(\langle i-1iAB \rangle \langle i+2iAj+1\rangle - \langle i-1iAj+1\rangle \langle i+2iAB\rangle\right)\\
&&= \sigma_L(B)\sigma_L(j+1)\langle i-1iAi+2\rangle \langle iABj+1\rangle >0 
\eeas
and
\beas
&&\langle i-1iAk \rangle \langle i+2iAk+1\rangle - \langle i-1iAk+1\rangle \langle i+2iAk\rangle \\
&&=\langle i-1iAi+2\rangle \langle iAkk+1\rangle >0 
\eeas
It is easy to see that these hold in all regions $(\mathcal{L}_i, \mathcal{R}_i)$.
As shown in Appendix[\ref{sec:app1}], we can conclude that the relation between $k_L$ and $k'_L$ depends only on the signs of first and last terms which are encoded in the matrix below.
\bea
\label{eq:signmatrix}
M &&= \begin{pmatrix}
\text{sign}(\langle i-1iAB\rangle) & \text{sign}(\langle i-2i-1iA\rangle) (-1)^{k_L}\\
\text{sign}(\langle i+2iAB\rangle) & \text{sign}(\langle i-2iAi+2\rangle) (-1)^{k_L}
\end{pmatrix} \\
&&\nonumber\equiv \begin{pmatrix}
+ & +\\
\text{sign}(\langle iAaa+1\rangle)\text{sign}(\langle Ai+1aa+1\rangle) & \text{sign}(\langle i-2ii+1i+2\rangle)
\end{pmatrix}
\eea
The relation between $k_L$ and $k'_L$ is tabulated below.\\
\begin{table}[htb!]
\begin{center}
\begin{tabular}{|c|c|c|c|}
\hline
$\text{sign}(\langle iAaa+1\rangle)$ & $\text{sign}(\langle Ai+1aa+1\rangle)$ & $\text{sign}(\langle i-2ii+1i+2\rangle)$ & $k_L-k'_L$\\
\hline
+&+&+&0\\
\hline
+&+&-&1\\
\hline
+&-&+&-1\\
\hline
+&-&-&0\\
\hline
-&+&+&-1\\
\hline
-&+&-&0\\
\hline
-&-&+&0\\
\hline
-&-&-&1\\
\hline
\end{tabular}
\end{center}
\caption{Relation between $k_L$ and $k'_L$ determined according to Appendix [\ref{sec:app1}]}
\end{table}
\\It is helpful to label all the allowed flip patterns of $S^\text{tree}$ as $S^\text{tree}_{ab}$ where $a$ and $b$ are the signs of $\langle ii+1i+2j\rangle$ and $\langle ii+1i+2j+1\rangle$ respectively. The different possibilities are shown below.
\beas
&&S^\text{tree}: \left\lbrace \langle i i+1 i+2 i+3 \rangle, \dots \langle i i+1 i+2 j \rangle \vline \langle i i+1 i+2 j+1\rangle, \dots, \langle i i+1 i+2 i-1\rangle (-1)^{k-1} \right\rbrace\\
&&S^\text{tree}_{++}: \left\lbrace + \hspace{2.5cm} k_1  \hspace{2.45cm} + \,\vline\, + \hspace{2.5cm} k_2 \hspace{3cm} (-1)^{k} \right\rbrace\\
&&S^\text{tree}_{+-}: \left\lbrace + \hspace{2.5cm} k_1  \hspace{2.45cm} + \,\vline\, - \hspace{2.5cm} k_2 \hspace{3cm} (-1)^{k} \right\rbrace\\
&&S^\text{tree}_{-+}: \left\lbrace + \hspace{2.5cm} k_1  \hspace{2.45cm} - \,\vline\, + \hspace{2.5cm} k_2 \hspace{3cm} (-1)^{k}  \right\rbrace\\
&&S^\text{tree}_{--}: \left\lbrace + \hspace{2.5cm} k_1  \hspace{2.45cm} - \,\vline\, - \hspace{2.5cm} k_2 \hspace{3cm} (-1)^{k} \right\rbrace\\
\eeas
For each configuration, $S^{\text{tree}}_{ab}$, the sequences $S_\mathcal{L}$ and $S_\mathcal{R}$ have the following signs depending on the region $(\mathcal{L}_i, \mathcal{R}_i)$.
\newpage
\begin{table}[htb!]
\begin{center}
\begin{tabular}{|c|c|c|c|c|}
\hline
$S_{++}^{\text{tree}}$&$\mathcal{R}_1$&$\mathcal{R}_2$&$\mathcal{R}_3$&$\mathcal{R}_4$\\
\hline
$\mathcal{L}_1$&$(k_2+1,k_1)$&$(k_2-1,k_1)$&$(k_2+1,k_1+1)$&$(k_2-1,k_1+1)$\\
\hline
$\mathcal{L}_2$&$ (k_2-1,k_1)$&$(k_2+1,k_1)$&$(k_2-1,k_1+1)$&$(k_2+1,k_1+1)$\\
\hline
$\mathcal{L}_3$&$(k_2,k_1)$&$(k_2,k_1)$&$(k_2,k_1+1)$&$(k_2,k_1+1)$\\
\hline
$\mathcal{L}_4$&$(k_2,k_1)$&$(k_2,k_1)$&$(k_2,k_1+1)$&$(k_2,k_1+1)$\\
\hline
\end{tabular}
\end{center}
\caption{$(k_L, k_R)$ in all regions for the configuration $S_{++}$ }
\end{table}
For the configuration $S_{++}$, we have $k_1+k_2=k$. Thus regions which satisfies $k_L+k_R=k$ are $(\mathcal{L}_1, \mathcal{R}_4),(\mathcal{L}_2, \mathcal{R}_3),(\mathcal{L}_3, \mathcal{R}_1),(\mathcal{L}_3, \mathcal{R}_2),(\mathcal{L}_4, \mathcal{R}_1),(\mathcal{L}_4, \mathcal{R}_2)$.
\begin{table}[htb!]
\begin{center}
\begin{tabular}{|c|c|c|c|c|}
\hline
$S_{+-}^{\text{tree}}$&$\mathcal{R}_1$&$\mathcal{R}_2$&$\mathcal{R}_3$&$\mathcal{R}_4$\\
\hline
$\mathcal{L}_1$&$(k_2,k_1)$&$(k_2,k_1)$&$(k_2,k_1+1)$&$(k_2,k_1+1)$\\
\hline
$\mathcal{L}_2$&$ (k_2,k_1)$&$(k_2,k_1)$&$(k_2,k_1+1)$&$(k_2,k_1+1)$\\
\hline
$\mathcal{L}_3$&$(k_2+1,k_1)$&$(k_2-1,k_1)$&$(k_2+1,k_1+1)$&$(k_2-1,k_1+1)$\\
\hline
$\mathcal{L}_4$&$(k_2-1,k_1)$&$(k_2+1,k_1)$&$(k_2-1,k_1+1)$&$(k_2+1,k_1+1)$\\
\hline
\end{tabular}
\end{center}
\caption{$(k_L, k_R)$ in all regions for the configuration $S_{+-}$ }
\end{table}
For the configuration $S_{+-}$, we have $k_1+k_2=k-1$. Thus regions which satisfies $k_L+k_R=k$ are $(\mathcal{L}_1, \mathcal{R}_3),(\mathcal{L}_1, \mathcal{R}_4),(\mathcal{L}_2, \mathcal{R}_3),(\mathcal{L}_2, \mathcal{R}_4),(\mathcal{L}_3, \mathcal{R}_1),(\mathcal{L}_4, \mathcal{R}_2)$.
\begin{table}[htb!]
\begin{center}
\begin{tabular}{|c|c|c|c|c|}
\hline
$S_{-+}^{\text{tree}}$&$\mathcal{R}_1$&$\mathcal{R}_2$&$\mathcal{R}_3$&$\mathcal{R}_4$\\
\hline
$\mathcal{L}_1$&$(k_2+1,k_1+1)$&$(k_2-1,k_1+1)$&$(k_2+1,k_1)$&$(k_2-1,k_1)$\\
\hline
$\mathcal{L}_2$&$ (k_2-1,k_1+1)$&$(k_2+1,k_1+1)$&$(k_2-1,k_1)$&$(k_2+1,k_1)$\\
\hline
$\mathcal{L}_3$&$(k_2,k_1+1)$&$(k_2,k_1+1)$&$(k_2,k_1)$&$(k_2,k_1)$\\
\hline
$\mathcal{L}_4$&$(k_2,k_1+1)$&$(k_2,k_1+1)$&$(k_2,k_1)$&$(k_2,k_1)$\\
\hline
\end{tabular}
\end{center}
\caption{$(k_L, k_R)$ in all regions for the configuration $S_{-+}$ }
\end{table}
\\ For the configuration $S_{-+}$, we have $k_1+k_2=k-1$. Thus regions which satisfies $k_L+k_R=k$ are $(\mathcal{L}_1, \mathcal{R}_3),(\mathcal{L}_2, \mathcal{R}_4),(\mathcal{L}_3, \mathcal{R}_1),(\mathcal{L}_3, \mathcal{R}_2),(\mathcal{L}_4, \mathcal{R}_1),(\mathcal{L}_4, \mathcal{R}_2)$.
\begin{table}[htb!]
\begin{center}
\begin{tabular}{|c|c|c|c|c|}
\hline
$S_{--}^{\text{tree}}$&$\mathcal{R}_1$&$\mathcal{R}_2$&$\mathcal{R}_3$&$\mathcal{R}_4$\\
\hline
$\mathcal{L}_1$&$(k_2,k_1+1)$&$(k_2,k_1+1)$&$(k_2,k_1)$&$(k_2,k_1)$\\
\hline
$\mathcal{L}_2$&$ (k_2,k_1+1)$&$(k_2,k_1+1)$&$(k_2,k_1)$&$(k_2,k_1)$\\
\hline
$\mathcal{L}_3$&$(k_2+1,k_1+1)$&$(k_2-1,k_1+1)$&$(k_2+1,k_1)$&$(k_2-1,k_1)$\\
\hline
$\mathcal{L}_4$&$(k_2-1,k_1+1)$&$(k_2+1,k_1+1)$&$(k_2-1,k_1)$&$(k_2+1,k_1)$\\
\hline
\end{tabular}
\end{center}
\caption{$(k_L, k_R)$ in all regions for the configuration $S_{--}$ }
\end{table}
\\For the configuration $S_{--}$, we have $k_1+k_2=k$. Thus regions which satisfies $k_L+k_R=k$ are $(\mathcal{L}_1, \mathcal{R}_3),(\mathcal{L}_1, \mathcal{R}_4),(\mathcal{L}_2, \mathcal{R}_3),(\mathcal{L}_2, \mathcal{R}_4),(\mathcal{L}_3, \mathcal{R}_2),(\mathcal{L}_4, \mathcal{R}_1)$.
\\

We see that for every configuration $S_{ab}^{\text{tree}}$, there are regions $(\mathcal{L}_i, \mathcal{R}_i)$ that satisfy $k_L+k_R = k$. Thus every configuration in the original amplituhedron can be covered by these regions consistent with the expected factorization. The remaining regions exist because they are related to amplitudes via inverse soft factors and have identical canonical forms. However, these are not necessary to cover all regions of the original amplituhedron.  
\subsection{Factorization of loop level data}
At loop level, we need to show that each loop $(AB)_a$ belongs either to the left or the right amplituhedron. The relevant sequences are (denoting $ \langle (AB)_a ij \rangle$ as $\langle ij \rangle$)
\beas
&&S^{\text{loop}}_R = \left\lbrace \langle i+1i+2 \rangle \sigma_R(i+2), \dots, \langle i+1j\rangle \sigma_R(j), \langle i+1B\rangle \sigma_R(B), (-1)^{k_r-1} \langle i+1A\rangle  \sigma_R(A)\right\rbrace\\
&&S^{\text{loop}}_L = \left\lbrace \langle iA \rangle \sigma_L(A), \langle iB \rangle \sigma_L(B), \langle ij+1\rangle \sigma_L(j+1), \dots, \langle ii-1\rangle \sigma_L(i-1) (-1)^{k_l-1}\right\rbrace
\eeas
Similar to before, it will be convenient to introduce the sequence $S_L^{'\text{loop}}$  
\beas
S^{'\text{loop}}_L = \left\lbrace \langle i+1A \rangle \sigma_L(A), \langle i+1B \rangle \sigma_L(B), \langle i+1j+1\rangle \sigma_L(j+1), \dots, \langle i+1i-1\rangle \sigma_L(i-1) (-1)^{k_l-1}\right\rbrace
\eeas
Let the number of flips in $S^{\text{loop}}_R$ and $S^{\text{loop}}_L$ be $k_r$ and $k_l$ respectively. These are {\it not} $k_R$ and $k_L$, which are the number of flips in the tree level sequences $S^{\text{tree}}_R$ and $S^{\text{tree}}_L$ respectively. The flip patterns of S$_{\text{tree}}$ can be organized as follows.
\beas
&&S^{\text{loop}}:\left\lbrace \langle i+1i+2\rangle, \langle i+1i+3\rangle, \dots, \langle i+1j\rangle \vline \langle i+1j+1\rangle, \dots, \langle i+1i\rangle (-1)^{k-1} \right\rbrace\\
&&S^{\text{loop}}_{++}:\left\lbrace\hspace{1cm}+ \hspace{2cm} k_1 \hspace{2cm} +\hspace{.45cm}\vline \hspace{.8cm}+ \hspace{1.2cm} k_2 \hspace{1.6cm} (-1)^k \hspace{.3cm}\right\rbrace\\
&&S^{\text{loop}}_{+-}:\left\lbrace\hspace{1cm}+ \hspace{2cm} k_1 \hspace{2cm} +\hspace{.45cm}\vline \hspace{.8cm}- \hspace{1.2cm} k_2 \hspace{1.6cm} (-1)^k \hspace{.3cm}\right\rbrace\\
&&S^{\text{loop}}_{-+}:\left\lbrace\hspace{1cm}+ \hspace{2cm} k_1 \hspace{2cm} -\hspace{.45cm}\vline \hspace{.8cm}+ \hspace{1.2cm} k_2 \hspace{1.6cm} (-1)^k \hspace{.3cm}\right\rbrace\\
&&S^{\text{loop}}_{--}:\left\lbrace\hspace{1cm}+ \hspace{2cm} k_1 \hspace{2cm} -\hspace{.45cm}\vline \hspace{.8cm}- \hspace{1.2cm} k_2 \hspace{1.6cm} (-1)^k \hspace{.3cm}\right\rbrace
\eeas
We showed in the previous section that on the unitarity cut, the external data factorizes such that $k_L+k_R = k$ with $k_L, k_R \in \left\lbrace 0, \dots k \right\rbrace$. It is trivially true that each loop belongs either to the left or the right amplituhedron. We must show that if a loop $(AB)_a$ belongs to the left amplituhedron, then it cannot belong to the right amplituhedron. First, note that in each configuration, we will have $k_l = k_2+l$ and $k_r = k_1+r$ with $r,l = 1$ or $2$. Now suppose that $(AB)_a$ belongs to both the left and right amplituhedra. Then we must have $k_l = k_L + 2$ and $k_r = k_R+2$. Expressing $k_l$ and $k_r$ in terms of $k_1, l, k_2$ and $r$, and using $k_L+k_R=k$, we get
\beas
l+r = \begin{cases}
4 \text{ if } k_1+k_2 = k\\
5 \text{ if } k_1+k_2 = k-1
\end{cases}
\eeas
Clearly, $l+r = 5$ is impossible since $l, r = 1$ or $ 2$. We just need to show that $l+r = 4$ is impossible. Note that this is possible only if $l=r =2$. In this case, the following hold true.
\beas
&&\left(\langle i+1j\rangle \sigma_R(j), \langle i+1B\rangle \sigma_R(B), (-1)^{k_r} \right) \sim (-1)^{k_r}(+,-,+)
\eeas
\beas
&&\left(\langle i+1A\rangle \sigma_L(A), \langle i+1B\rangle \sigma_L(B), \langle i+1j+1\rangle \sigma_L(j+1) \right) \\
=&&\left(-\sigma_L(A)\sigma_R(A), -(-1)^{k_r} \sigma_R(B)\sigma_L(B), \langle i+1j+1\rangle \sigma_L(j+1)\right)\\
=&&(+,-,+) \text{ or } (-,+,-)
\eeas 
In all these cases, we must have $\sigma_R(A)\sigma_R(B)\sigma_L(A)\sigma_L(B)(-1)^{k_R} <0$. It is easy to verify from Section [\ref{sec:leftrightamp}] that this is always false. Thus each loop belongs solely to the left or the right ampltuhedron.
\subsection{Mutual positivity}
\label{sec:higherkmutual}
To complete the proof of factorization, we need to show that the mutual positivity between a loop in $\mathcal{A}^{\mathcal{L}}_{n_1,k_L,L_1}$ and one in $\mathcal{A}^{\mathcal{R}}_{n_2,k_R,L_2}$ is automatically satisfied. This is easier to see while working with $(k+2)$ dimensional data. We can re-write all the four brackets using $\mathcal{Z}'s$ and the $k-$plane $Y$ as described in Section [\ref{sec:momtwist}]. For more details, see Section [7] of \cite{binarycode}. A loop in the left amplituhedron can be parametrized as a $k_L+2$ plane $Y^L_1 \dots Y^L_{k_L}A_aB_a$. 
\bea
\label{eq:leftloop}
&&Y^L_{\nu} = (-1)^{\nu -1} \sigma_L(A) \, A + \alpha_{\nu} \,\sigma_L(i_\nu) \,\mathcal{Z}_{i_\nu} + \beta_{\nu}\, \sigma_L(\nu+1)\, Z_{i_{\nu+1}}\\
&&\nonumber A_{a} = (-1)^{k_L+1} \sigma_L(A)\, A + \alpha_{k_L+1}\,\sigma_L(i_{k_L+1}) \mathcal{Z}_{i_{k_{L}+1}} + \beta_{i_{k_L+1}} \, \sigma_L(i_{k_L+1}+1) \,\mathcal{Z}_{i_{k_L+1}+1}\\
&&\nonumber B_{a} = (-1)^{k_L+2} \sigma_L(A)\, A + \alpha_{k_L+2}\,\sigma_L(i_{k_L+2})\, \mathcal{Z}_{i_{k_L+2}} + \beta_{i_{k_L+2}} \,\sigma_L(i_{k_L+2}+1) \,\mathcal{Z}_{i_{k_L+2}+1}
\eea
with $\nu = \left\lbrace 1, \dots k_L \right\rbrace$, $ \mathcal{Z}_{i_\nu} \in \left\lbrace \mathcal{Z}_1, \dots \mathcal{Z}_i, A, B, \mathcal{Z}_{j+1}, \mathcal{Z}_n\right\rbrace $ and $i_1<i_2<\dots i_{K_L}+2 $.\\

Similarly, a loop in the right amplituhedron can be thought of as a $k_R+2$ plane $Y^R_1 \dots Y^R_{k_R}A_bB_b$ and parametrized as 
\bea
\label{eq:rightloop}
&&Y^R_{\mu} = (-1)^{\mu -1} \sigma_R(A) \, A + \alpha_{\mu} \,\sigma_R(i_\mu) \,Z_{i_\mu} + \beta_{\mu}\, \sigma_R(\mu+1)\, \mathcal{Z}_{i_{\mu+1}}\\
&&\nonumber A_{b} = (-1)^{k_R+1} \sigma_R(A)\, A + \alpha_{k_R+1}\,\sigma_R(i_{k_R+1}) \mathcal{Z}_{i_{k_R+1}} + \beta_{i_{k_R+1}} \, \sigma_R(i_{k_R+1}+1) \,\mathcal{Z}_{i_{k_R+1}+1}\\
&&\nonumber B_{b} = (-1)^{k_R+2} \sigma_R(A)\, A + \alpha_{k_R+2}\,\sigma_R(i_{k_R+2})\, \mathcal{Z}_{i_{k_R+2}} + \beta_{i_{k_R+2}} \,\sigma_R(i_{k_R+2}+1) \,\mathcal{Z}_{i_{k_R+2}+1}
\eea
with $\mu \in \left\lbrace 1, \dots k_R \right\rbrace $, $\mathcal{Z}_{i_{\mu}} \in \left\lbrace A, \mathcal{Z}_{i+1}, \dots, \mathcal{Z}_j, B\right\rbrace$  and with $j_1<j_2<\dots j_{k_R+2}$.\\

This reduces the mutual positivity condition $\langle Y^L(AB)_aY^R(AB)_b \rangle > 0$ to a condition involving $k+4$ brackets of the form $\langle ijklm\rangle$. It is easy to see that with positive $k+4$ dimensional data $(\langle i_1 \dots i_{k+4}\rangle$ when $i_1<i_2< \dots i_{k+4})$, mutual positivity is guaranteed. The signs $\sigma_L(k)$ and $\sigma_R(k)$ are crucial in making this work.  
\section{Conclusions}
We have shown that unitarity can be an emergent feature. The positivity of the geometry inevitably leads to amplitudes identical to those derived from a unitary quantum field theory. This lends further support for the conjecture that the amplituhedron computes all the amplitudes of $\mathcal{N}=4$ SYM. It also suggests that the notion of positivity is more fundamental than those of unitarity and locality which are the cornerstones of the traditional framework of quantum field theory.
\acknowledgments
We would like to thank Nima Arkani-Hamed and Jaroslav Trnka for useful discussions and comments about the manuscript.
\appendix
\section{Restricting flip patterns}
\label{sec:app1}
Consider a pair of sequences
$\left\lbrace a_1, \dots, a_n \right\rbrace$ and $\left\lbrace b_1, \dots, b_n \right\rbrace$ which have an equal number of terms. Further suppose that they are connected by the Schouten identity and satisfy a postivity condition, i.e. there exists a relation $a_ib_{i+1}-a_{i+1}b_i = ab >0$. We will show that the number of sign flips in these sequences, $k_1$ and $k_2$ respectively, are related and that the relation depends only on the signs of $a_1, a_n, b_1$ and $b_n$. \\

\noindent Firstly, we note that the positivity forces each block in the pair of sequences
$\begin{pmatrix}
a_i & a_{i+1}\\
b_i & b_{i+1}
\end{pmatrix}$
to take one of the following forms.
\beas
&&\textbf{Type 1:}\begin{pmatrix}
+ && + \\
+ && +
\end{pmatrix}, \begin{pmatrix}
+ && + \\
- && -
\end{pmatrix} ,
\begin{pmatrix}
- && - \\
+ && +
\end{pmatrix}, \begin{pmatrix}
- && - \\
- && -
\end{pmatrix} \\
&&\textbf{Type 2:}\begin{pmatrix}
+ && - \\
- && +
\end{pmatrix}, \begin{pmatrix}
+ && - \\
+ && -
\end{pmatrix},
\begin{pmatrix}
- && + \\
+ && -
\end{pmatrix}, 
\begin{pmatrix}
- && + \\
- && +
\end{pmatrix}\\
&&\textbf{Type 3:}\begin{pmatrix}
+ && + \\
- && +
\end{pmatrix},
\begin{pmatrix}
- && - \\
+ && -
\end{pmatrix}\\
&&\textbf{Type 4:}
\begin{pmatrix}
+ && - \\
+ && +
\end{pmatrix}
\begin{pmatrix}
- && + \\
- && -
\end{pmatrix}, 
\eeas
Blocks of type 1 and 2 leave $k_1-k_2$ fixed. A block of type 3 changes $k_1-k_2$ by $-1$ and a block of type 4 changes it by $1$. Two consecutive blocks of type 3 or 4 are prohibited and a block of type 4 must follow  a block of type 3 before the sign of the bottom sequence can be flipped without flipping the sign of the top. Thus, if we know the signs of $a_1, a_n, b_1$ and $b_n$, we can determine $k_1-k_2$.  	
We can list the possibilities by the matrices 
$\begin{pmatrix}
s(a_1) & s(a_n)\\
s(b_1) & s(b_n)
\end{pmatrix}$
where $s(x)$ is the sign of $x$.
\begin{itemize}
\item $k_1 = k_2$
\beas
\begin{pmatrix}
+ & +\\
+ & +
\end{pmatrix} 
\begin{pmatrix}
+ & -\\
+ & -
\end{pmatrix}
\begin{pmatrix}
+ & +\\
- & -
\end{pmatrix}
\begin{pmatrix}
+ & -\\
- & +
\end{pmatrix}
\begin{pmatrix}
- & -\\
+ & +
\end{pmatrix}
\begin{pmatrix}
- & +\\
+ & -
\end{pmatrix}
\begin{pmatrix}
- & -\\
- & -
\end{pmatrix}
\begin{pmatrix}
- & +\\
- & +
\end{pmatrix}
\eeas
\item $k_1 = k_2+1$
\beas
\begin{pmatrix}
+ & -\\
+ & +
\end{pmatrix}
\begin{pmatrix}
+ & +\\
+ & -
\end{pmatrix}
\begin{pmatrix}
- & +\\
- & -
\end{pmatrix}
\begin{pmatrix}
- & -\\
- & +
\end{pmatrix}
\eeas
\item $k_1=k_2-1$
\beas
\begin{pmatrix}
- & +\\
+ & +
\end{pmatrix}
\begin{pmatrix}
+ & +\\
- & +
\end{pmatrix}
\begin{pmatrix}
+ & -\\
- & -
\end{pmatrix}
\begin{pmatrix}
- & -\\
+ & -
\end{pmatrix}
\eeas
\end{itemize}

\bibliographystyle{unsrt}
\bibliography{Unitarity}

\end{document}